\documentclass[reprint,superscriptaddress,preprintnumbers,longbibliography,
amsmath,amssymb,aps,superscriptaddress,floatfix,tightenlines,pra,
]{revtex4-2}
\usepackage{subfigure}
\usepackage{mathrsfs}
\usepackage[T1]{fontenc}
\usepackage{graphicx}
\usepackage{dcolumn}
\usepackage{bm}
\usepackage{color}
\usepackage{times}
\usepackage[colorlinks=true,
citecolor=magenta,
linkcolor=magenta,
anchorcolor=black,
urlcolor=magenta]{hyperref}
\usepackage{siunitx}
\usepackage{overpic}
\usepackage{braket}
\usepackage{changes}
\definechangesauthor[name={syh},color=blue]{syh}

\begin{document}
	\title{Interplay between disorder and topology in Thouless pumping on a superconducting quantum processor}
	
	\author{Yu Liu}
	\thanks{These authors contributed equally to this work.}
	\affiliation{Institute of Physics, Chinese Academy of Sciences, Beijing 100190, China}
	\affiliation{School of Physical Sciences, University of Chinese Academy of Sciences, Beijing 100049, China}
	
	\author{Yu-Ran Zhang}
	\thanks{These authors contributed equally to this work.}
	\affiliation{School of Physics and Optoelectronics, South China University of Technology, Guangzhou 510640, China}
	
	\author{Yun-Hao Shi}
	\affiliation{Institute of Physics, Chinese Academy of Sciences, Beijing 100190, China}
	
	\author{Tao Liu}
	\affiliation{School of Physics and Optoelectronics, South China University of Technology, Guangzhou 510640, China}
	
	\author{Congwei Lu}
	\affiliation{Department of Physics, Applied Optics Beijing Area Major Laboratory, Beijing Normal University, Beijing 100875, China}
	
	\author{Yong-Yi Wang}
	\affiliation{Institute of Physics, Chinese Academy of Sciences, Beijing 100190, China}
	\affiliation{School of Physical Sciences, University of Chinese Academy of Sciences, Beijing 100049, China}
	
	\author{Hao Li}
	\affiliation{Institute of Physics, Chinese Academy of Sciences, Beijing 100190, China}
	
	\author{Tian-Ming Li}
	\affiliation{Institute of Physics, Chinese Academy of Sciences, Beijing 100190, China}
	\affiliation{School of Physical Sciences, University of Chinese Academy of Sciences, Beijing 100049, China}
	
	\author{Cheng-Lin Deng}
	\affiliation{Institute of Physics, Chinese Academy of Sciences, Beijing 100190, China}
	\affiliation{School of Physical Sciences, University of Chinese Academy of Sciences, Beijing 100049, China}
	
	\author{Si-Yun Zhou}
	\affiliation{Institute of Physics, Chinese Academy of Sciences, Beijing 100190, China}
	\affiliation{School of Physical Sciences, University of Chinese Academy of Sciences, Beijing 100049, China}
	
	\author{Tong Liu}
	\affiliation{Institute of Physics, Chinese Academy of Sciences, Beijing 100190, China}
	
	\author{Jia-Chi Zhang}
	\affiliation{Institute of Physics, Chinese Academy of Sciences, Beijing 100190, China}
	\affiliation{School of Physical Sciences, University of Chinese Academy of Sciences, Beijing 100049, China}
	
	\author{Gui-Han Liang}
	\affiliation{Institute of Physics, Chinese Academy of Sciences, Beijing 100190, China}
	\affiliation{School of Physical Sciences, University of Chinese Academy of Sciences, Beijing 100049, China}
	
	\author{Zheng-Yang Mei}
	\affiliation{Institute of Physics, Chinese Academy of Sciences, Beijing 100190, China}
	\affiliation{School of Physical Sciences, University of Chinese Academy of Sciences, Beijing 100049, China}
	
	\author{Wei-Guo Ma}
	\affiliation{Institute of Physics, Chinese Academy of Sciences, Beijing 100190, China}
	\affiliation{School of Physical Sciences, University of Chinese Academy of Sciences, Beijing 100049, China}
	
	\author{Hao-Tian Liu}
	\affiliation{Institute of Physics, Chinese Academy of Sciences, Beijing 100190, China}
	\affiliation{School of Physical Sciences, University of Chinese Academy of Sciences, Beijing 100049, China}
	
	\author{Zheng-He Liu}
	\affiliation{Institute of Physics, Chinese Academy of Sciences, Beijing 100190, China}
	\affiliation{School of Physical Sciences, University of Chinese Academy of Sciences, Beijing 100049, China}
	
	\author{Chi-Tong Chen}
	\affiliation{Quantum Science Center for Guangdong-Hong Kong-Macao Greater Bay Area, 518045 Shenzhen, Guangdong, China}
	
	\author{Kaixuan Huang}
	\affiliation{Beijing Academy of Quantum Information Sciences, Beijing 100193, China}
	
	\author{Xiaohui Song}
	\affiliation{Institute of Physics, Chinese Academy of Sciences, Beijing 100190, China}
	
	\author{S. P. Zhao}
	\affiliation{Institute of Physics, Chinese Academy of Sciences, Beijing 100190, China}
	\affiliation{School of Physical Sciences, University of Chinese Academy of Sciences, Beijing 100049, China}
	\affiliation{Songshan Lake  Materials Laboratory, Dongguan, Guangdong 523808, China}
	
	\author{Ye Tian}
	\affiliation{Institute of Physics, Chinese Academy of Sciences, Beijing 100190, China}
	
	\author{Zhongcheng Xiang}
	\email{zcxiang@iphy.ac.cn}
	\affiliation{Institute of Physics, Chinese Academy of Sciences, Beijing 100190, China}
	\affiliation{School of Physical Sciences, University of Chinese Academy of Sciences, Beijing 100049, China}
	\affiliation{Hefei National Laboratory, Hefei 230088, China}
	
	\author{Dongning Zheng}
	\affiliation{Institute of Physics, Chinese Academy of Sciences, Beijing 100190, China}
	\affiliation{School of Physical Sciences, University of Chinese Academy of Sciences, Beijing 100049, China}
	\affiliation{Songshan Lake Materials Laboratory, Dongguan, Guangdong 523808, China}
	\affiliation{CAS Center for Excellence in Topological Quantum Computation, UCAS, Beijing 100049, China}
	\affiliation{Hefei National Laboratory, Hefei 230088, China}
	
	\author{Franco Nori}
	\affiliation{Theoretical Quantum Physics Laboratory, Cluster for Pioneering Research, RIKEN, Wako-shi, Saitama 351-0198, Japan}
	\affiliation{Center for Quantum Computing, RIKEN, Wako-shi, Saitama 351-0198, Japan}
	\affiliation{Physics Department, University of Michigan, Ann Arbor, Michigan 48109-1040, USA}
	
	\author{Kai Xu}
	\email{kaixu@iphy.ac.cn}
	\affiliation{Institute of Physics, Chinese Academy of Sciences, Beijing 100190, China}
	\affiliation{School of Physical Sciences, University of Chinese Academy of Sciences, Beijing 100049, China}
	\affiliation{Beijing Academy of Quantum Information Sciences, Beijing 100193, China}
	\affiliation{Songshan Lake Materials Laboratory, Dongguan, Guangdong 523808, China}
	\affiliation{CAS Center for Excellence in Topological Quantum Computation, UCAS, Beijing 100049, China}
	\affiliation{Hefei National Laboratory, Hefei 230088, China}
	
	\author{Heng Fan}
	\email{hfan@iphy.ac.cn}
	\affiliation{Institute of Physics, Chinese Academy of Sciences, Beijing 100190, China}
	\affiliation{School of Physical Sciences, University of Chinese Academy of Sciences, Beijing 100049, China}
	\affiliation{Beijing Academy of Quantum Information Sciences, Beijing 100193, China}
	\affiliation{Songshan Lake Materials Laboratory, Dongguan, Guangdong 523808, China}
	\affiliation{CAS Center for Excellence in Topological Quantum Computation, UCAS, Beijing 100049, China}
	\affiliation{Hefei National Laboratory, Hefei 230088, China}
	
	\begin{abstract}
		Topological phases are robust against weak perturbations, but break down when
		disorder becomes sufficiently strong.
		However, moderate disorder can also induce topologically nontrivial phases.
		Thouless pumping, as a (1+1)D counterpart of the integer quantum Hall effect, is one of the simplest manifestations of topology. 
		Here, we report experimental observations of the competition and interplay between Thouless pumping and disorder on
		a 41-qubit superconducting quantum processor. 
		We improve a Floquet engineering technique to realize cycles of adiabatic pumping by simultaneously varying the on-site potentials and the hopping couplings. We
		demonstrate Thouless pumping in the presence of disorder and show its breakdown as the strength of disorder increases. Moreover, we observe two types of topological pumping that are induced by on-site potential disorder and hopping disorder, respectively. In particular, an intrinsic topological pump that is induced by quasi-periodic hopping disorder has never been experimentally realized before. Our highly controllable system provides a valuable quantum simulating platform for studying various aspects of topological physics in the presence of disorder.
	\end{abstract}
	\date{\today}
	
	\maketitle
	Topology versus disorder provides a diverse landscape for exploration in modern condensed matter physics, ranging from the robustness of topological systems against weak disorder~\cite{moore_nature_2010} to the classification of symmetry-protected topological phases~\cite{qi_topological_2011}. One of the most significant class of topological systems is the Thouless pump~\cite{thouless_quantization_1983, niu_quantised_1984}, entailing transport of the quantized charge during an adiabatic cyclic evolution of the underlying Hamiltonian~\cite{niu_quantised_1984,citro_thouless_2023}. Thouless pumping, as a dynamical version of the integer quantum Hall effect (IQHE)~\cite{thouless_hall2d_1983},
	bridges the quantized conductance and the topological invariant,
	known as the Chern number of the occupied energy bands~\cite{xiao_berry_2010,moore_nature_2010}.
	Due to the universality of topological effects, the Thouless pump is not a specific phenomenon occurring in a certain system and is robust against perturbations~\cite{niu_quantised_1984,citro_thouless_2023}. These properties make topological pumps a promising platform for designing novel devices with unprecedented functionalities~\cite{citro_thouless_2023}.
	Thouless pumping has been experimentally demonstrated on
	different experimental platforms \cite{nakajima_topological_2016,lohse_thouless_2016,lohse_nature_2018,
		cheng_experimental_2020,kao_science_2021,jurgensen_nature_2021,
		mostaan_pump_2022,you_prl_2022,xiang_simulating_2022,dreon_nature_2022,
		tao_interaction-induced_2023,jurgensen_quantized_2023}.
	Especially, the competition and interplay between topology and disorder in a Thouless pump have been attracting growing attention in, e.g., ultra-cold atoms~\cite{nakajima_competition_2021, walter_quantization_2023}, photonic waveguides~\cite{cerjan_thouless_2020}, and mechanical metamaterials~\cite{grinberg_robust_2020}.
	These experiments not only demonstrate topological transitions with disorder, but also the breakdown of 
	quantized pumps due to localization caused by disorder~\cite{wauters_localization_2019,ippoliti_dimensional_2020}.
	
	\begin{figure*}[t]
		\centering
		\includegraphics[width=0.97\textwidth]{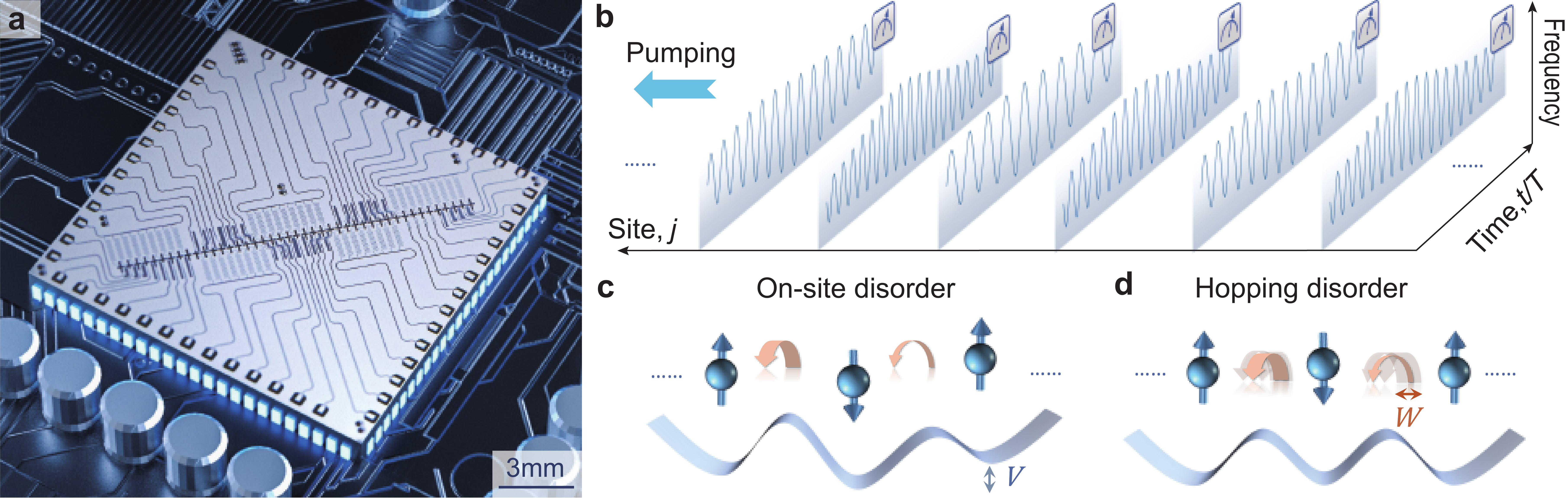}
		\caption{\textbf{Device and pulse sequences.}
			\textbf{a}, Optical micrograph of the 43-qubit superconducting chip.
			\textbf{b},  Schematic of Floquet engineering for the adiabatic cyclic evolution. Pulse sequences in one pumping period are illustrated.  The amplitude and the center shift of the Floquet engineering pulse on each qubit are subject to a cyclic modulation, corresponding to the cyclic variations of hopping couplings and on-site potentials, respectively.
			\textbf{c}, Schematic diagram of the Rice-Mele (RM) model with on-site potential disorder. The on-site potentials on qubits are staggered, with a random offset of disorder strength $V$. The orange curved arrows, representing the couplings, stagger with one large and one small, due to the staggered RM hopping Hamiltonian.
			\textbf{d}, Schematic diagram of the RM model in the presence of hopping disorder. The on-site potential is strictly periodic, while the disordered hopping coupling is modulated with disorder strength $W$. In the clean limit, the on-site potentials (hopping couplings), denoted by the blue spin (orange curved arrows), are staggered with one 
			up 
			(large) and one down (small) due to the staggered RM Hamiltonian.}
		\label{fig:fig1}
	\end{figure*}
	
	To exploit disorder rather than to eliminate it, we experimentally investigate Thouless pumping
	induced by disorder on a 41-qubit superconducting processor.
	Since it is challenging to precisely control the adiabatic cyclic evolution of a multi-qubit system with disorder, we employ a Floquet engineering technique~\cite{cai_observation_2019,zhao_probing_2022,shi2022observing} to realize Thouless pumping by simultaneously varying the on-site potentials and hopping strengths~\cite{supp_cite}. We experimentally demonstrate bulk topological pumping during different pumping trajectories in the clean limit.
	We also observe the breakdown of quantized pumping, when the strength of the random on-site potential disorder increases.
	For a topologically trivial double-loop pumping trajectory, we observe topological pumping 
	induced by the on-site disorder of a uniform random distribution. Moreover, we experimentally  demonstrate emergent topological pumping induced by quasi-periodic hopping disorder, which is related to the dynamic version of topological Anderson insulators (TAI)~\cite{li_tai_2009,groth_theory_2009,meier_observation_2018,stutzer_tai_2018,liu_topological_2020,wu_quantized_2022}. Our results will  inspire further investigations of topological phases 
	in the presence of disorder on quantum simulating platforms~\cite{georgescu_quantum_2014,you_atomic_2011,gu_microwave_2017,Cheng2023,satzinger_realizing_2021,semeghini_probing_2021,daley_practical_2022,zhang_digital_2022,li2023mapping}.
	\\
	\\
	\noindent \textbf{System and model}
	\\
	Our experiments are performed on a 1D superconducting processor, named \emph{Chuang-tzu}, consisting of 43 nearest-neighbor-coupled and frequency-tunable transmon qubits~\cite{shi2022observing}. In our experiments,  41 qubits ($Q_j$ with $j$ varying from 1 to 41) are used, and the system Hamiltonian is written as $\hat{H}_0=\sum_{j}[(g_{j,j+1}\hat{a}_j^{\dagger}\hat{a}_{j+1}+\mathrm{H.c.})+\omega_j\hat{n}_j]$,
	where $\hat{a}^{\dagger}$ ($\hat{a}$) denotes the hard-core bosonic creation (annihilation) operator~\cite{yan_strongly_2019}, $\hat{n}=\hat{a}^{\dagger}\hat{a}$ is the number operator, and $g_{j,j+1}$ is the nearest-neighbor (NN) hopping strength.

		\begin{figure*}[t]
		\centering
		\includegraphics[width=0.97\textwidth]{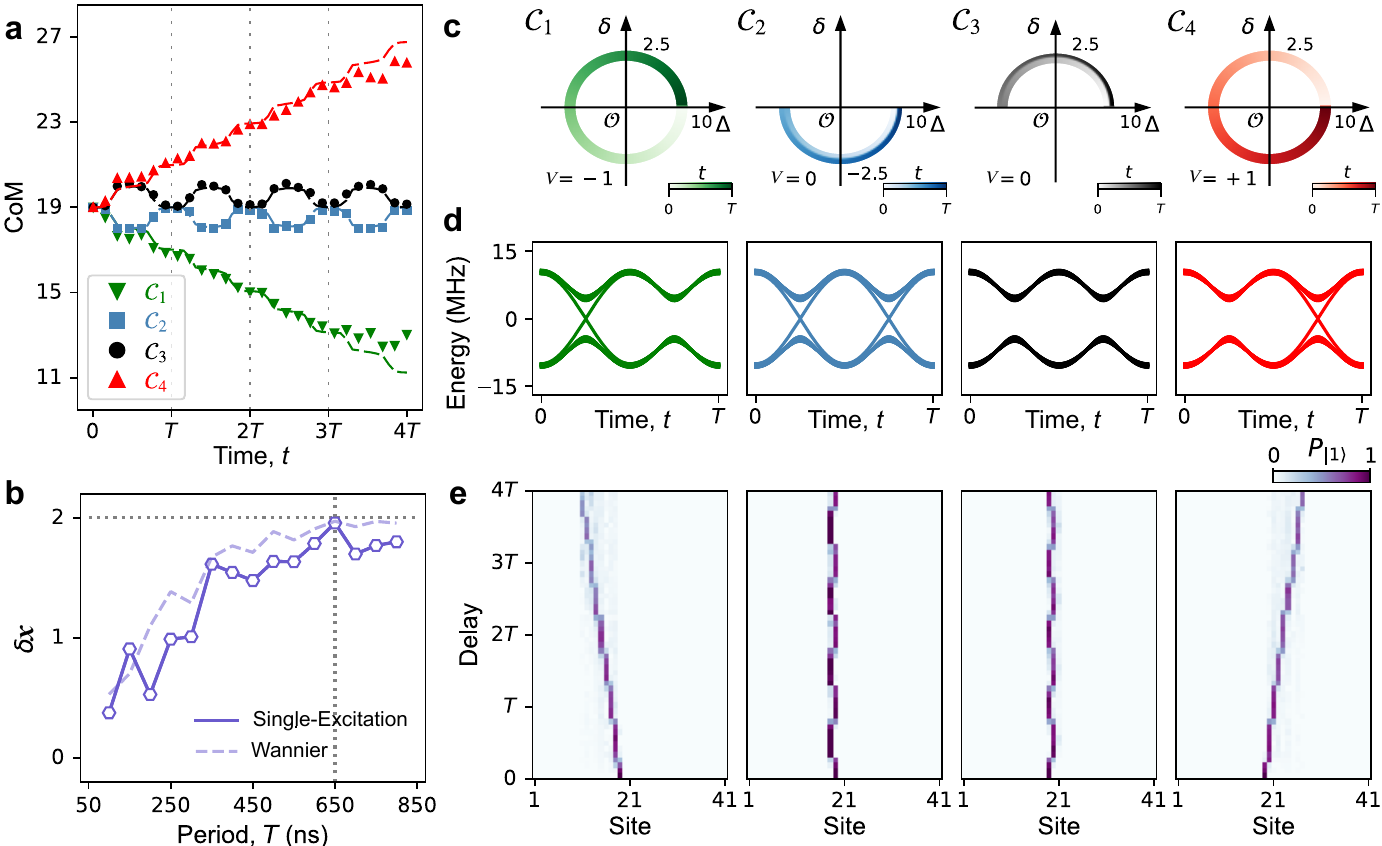}
		\caption{\textbf{Bulk topological pumping for different types of cycles initially with single-excitation states.} \textbf{a}, Displacements of the CoM for four different pumping trajectories $\mathcal{C}_1$--$~\!\mathcal{C}_4$ as illustrated in \textbf{c}.
			Dashed curves represent the numerical results. \textbf{b}, Displacement of CoM, $\delta x$, versus the pumping period $T$, for $\mathcal{C}_4$ initially with a single-excitation  state. When $T=650~\mathrm{ns}$, $\delta x$ reaches the maximum $1.95(6)$. The dashed curve shows the numerical results of $\delta x$ as a function of $T$, when the initial state is an exact Wannier state. \textbf{c}, Four different pumping trajectories $\mathcal{C}_1$--$~\!\mathcal{C}_4$ in the $\delta$--$\Delta$ plane.
			The trajectory $\mathcal{C}_4$ is set as $(\Delta, \delta)=(\Delta_0\cos 2\pi t/T, \delta_0\sin 2\pi t/T)$, with $\Delta_0/2\pi=10~\mathrm{MHz}$, $\delta_0/2\pi=2.5~\mathrm{MHz}$, and $J/2\pi=2~\mathrm{MHz}$.
			The trajectory $\mathcal{C}_3$ is set as $(\Delta, \delta)=(\Delta_0\cos 2\pi t/T, \delta_0|\sin 2\pi t/T|)$ with the same parameters $\Delta_0$, $\delta_0$, and $T$ as $\mathcal{C}_4$.
			The trajectory $\mathcal{C}_1$ ($\mathcal{C}_2$) is designed symmetrically flipped about the $\Delta$-axis with $\mathcal{C}_4$ ($\mathcal{C}_3$).
			The trajectory $\mathcal{C}_1$ and $\mathcal{C}_4$ correspond to the Chern numbers $\nu=\mp1$, respectively, and $\mathcal{C}_2$ and $\mathcal{C}_3$ lead to topologically trivial pumping. \textbf{d}, Instantaneous energy spectra of the bulk under open boundary conditions. \textbf{e}, Experimental data of the populations of all qubits during the adiabatic cyclic evolution within four periods.}
		\label{fig:fig2}
	\end{figure*}
	
	To experimentally demonstrate a disorder-induced pumping process,
	we simulate the tight-binding Rice-Mele (RM) model with on-site potential disorder or hopping disorder, of which the Hamiltonian can be expressed as~\cite{rice_elementary_1982}:
	\begin{align}
		\hat{H}_\textrm{RM}(t)=&\sum_{j=1}^{40} \{J+(-)^{j-1}[\delta(t)+W_j]\}(\hat{a}_j^{\dagger}\hat{a}_{j+1}+\mathrm{H.c.})\nonumber\\
		&+\sum_{j=1}^{41} (-)^{j-1}[\Delta(t)+V_j] \hat{n}_j.\label{RMM}
	\end{align}
	Here, $J\pm[\delta(t)+W_j]$ denote the NN hopping strengths with  disorder $W_j$, $\pm[\Delta(t)+V_j]$ denote the staggered on-site potential with disorder $V_j$, and $\Delta (t)$ and $\delta (t)$ are periodic with the period $T$. When $\Delta(t)=0$, the RM model reduces to the Su-Shrieffer-Heeger (SSH) model~\cite{ssh_1979} in the clean limit.
	Furthermore, to realize the adiabatic cyclic evolution of the RM Hamiltonian (\ref{RMM}), we develop a Floquet engineering technique to change the dynamical parameters $\delta(t)$ and $\Delta(t)$ adiabatically during a closed trajectory
	in a $\delta$--$\Delta$ space (Fig.~\ref{fig:fig1}b).
	More details are discussed in the Supplementary Materials~\cite{supp_cite}.
	We realize the pumping process with the cyclic modulations of both the amplitude and the center offset of the sine-like waves of Floquet engineering, corresponding to the cyclic variations of the hopping coupling and the on-site potential, respectively, where disorder is also carefully introduced.
	\\
	\\
	\noindent \textbf{Topological invariant and topological pumping}
	\\
	In the clean limit, the continuous RM pumping sequence is periodic in both spatial and temporal dimensions. Under periodic boundary conditions (PBCs), the Bloch wavefunction of the $n$-th energy band is defined in the $k$--$t$ Brillouin zone as $\ket{\psi_{k,n}(t)}=e^{\mathrm{i}kx}\ket{u_{n,k}(t)}$, and the Chern number is expressed as \cite{lohse_thouless_2016}
	\begin{equation}
		\nu_n=\frac{1}{2\pi}\int_{\text{FBZ}}\!\!\!\!\!\mathrm{d}k\;\int_0^T\!\!\!\mathrm{d}t\;\Omega_n(k, t),
	\end{equation}
	where $\Omega_n(k, t)=\mathrm{i}(\bra{\partial_tu_{n,k}}\partial_ku_{n,k}\rangle-\bra{\partial_ku_{n,k}}\partial_ku_{n,t}\rangle)$ denotes the Berry curvature, and $\text{FBZ}$ represents the first Brillouin zone. When the system is initially prepared as a Wannier state, filling the $n$-th band, $\nu_n$ relates to the displacement of the center-of-mass (CoM) per pumping cycle $\delta x$ as
	\begin{equation}
		\delta x=\nu_n d,
	\end{equation}
	with $d=2$ being the lattice constant~\cite{ke_multiparticle_2017}.
	
			\begin{figure*}[t]
		\centering
		\includegraphics[width=0.97\linewidth]{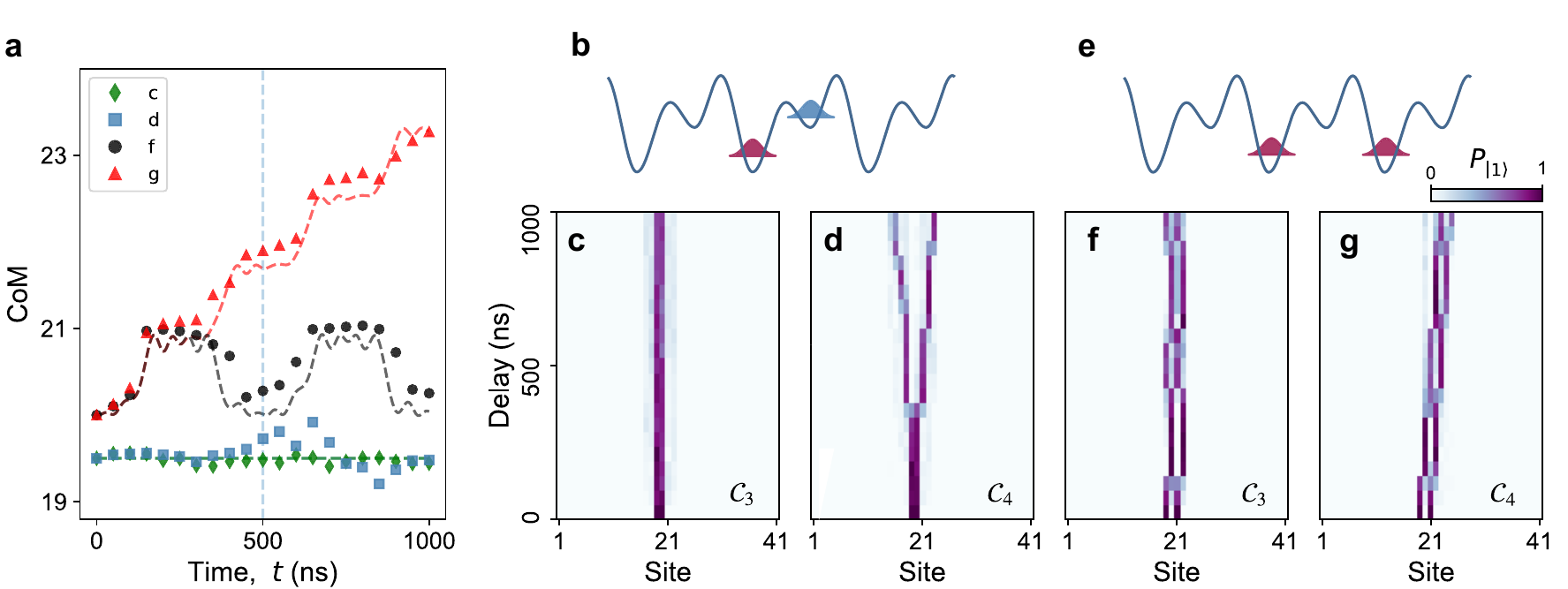}
		\caption{\textbf{Topological pumping for different types of cycles which initially have double-excitation states.} \textbf{a}, Displacements of the CoM for different pumping protocols as shown in \textbf{c}--\textbf{d},\textbf{f}--\textbf{g}, where the dashed curves represent the numerical results.
		\textbf{b}, Schematic diagram of lattice potentials with the initial excitions prepared at two nearest-neighbor sites, i.e., $Q_{19}$ and $Q_{20}$. \textbf{c} and \textbf{d}, Experimental data of the population of all qubits during the adiabatic cyclic evolution within two periods for the trajectory $\mathcal{C}_3$ and $\mathcal{C}_4$, respectively.
		\textbf{e}, Schematic diagram of lattice potential with the initial state prepared by exciting two next-neighbor sites, i.e., $Q_{19}$ and $Q_{21}$. Adiabatic time evolutions of the the populations of all qubits within two periods for the pumping trajectories $\mathcal{C}_3$ and $\mathcal{C}_4$ are shown in \textbf{f} and \textbf{g}. The evolution period for double-excitation pumping is $500$~ns.}
		\label{fig:figdouble}
	\end{figure*}
	
	%
	
	\begin{figure*}[t]
		\centering
		\includegraphics[width=0.97\linewidth]{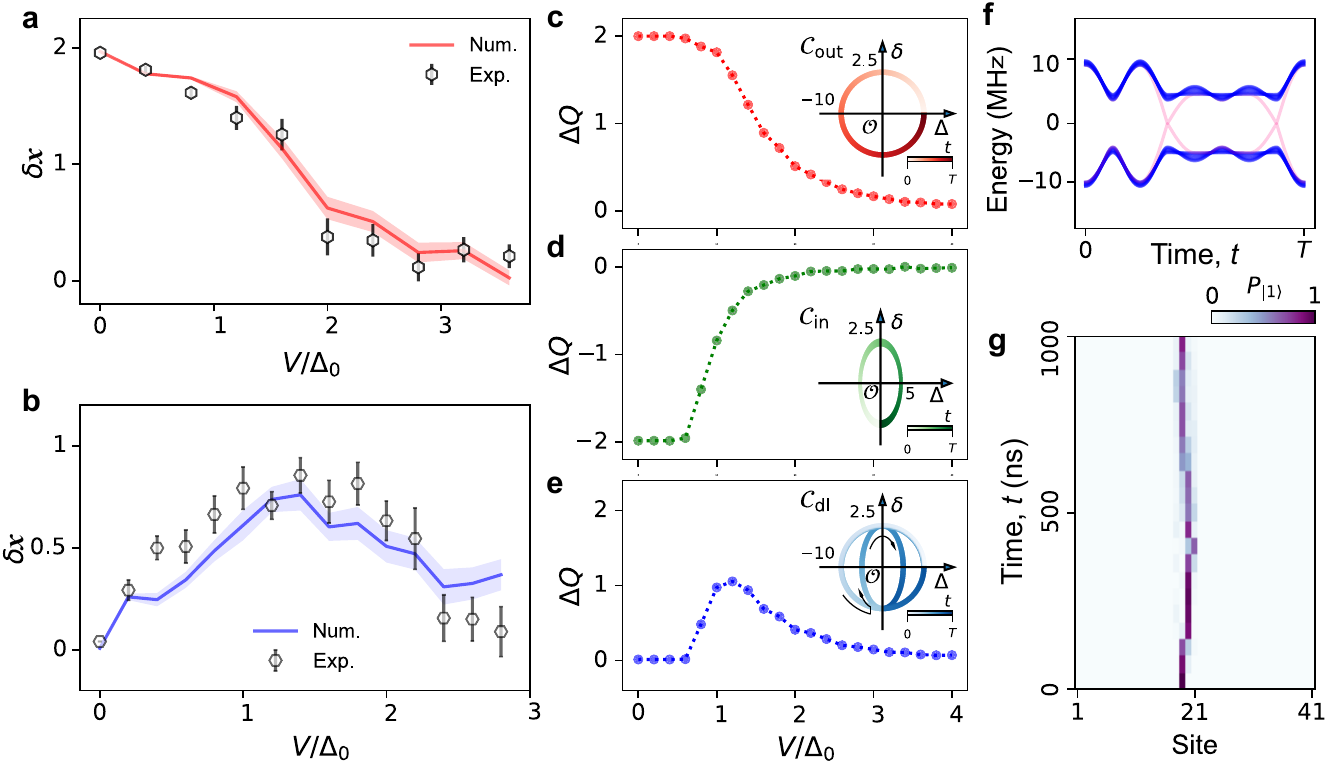}
		\caption{\textbf{Pumping with on-site disorder.} \textbf{a}, Experimental data of the displacement of the CoM $\delta x$ for 2 pumping periods against the on-site potential disorder strength $V$ divided by $\Delta_0/2\pi=10$~MHz during the pumping trajectory $\mathcal{C}_{\text{out}}$ as plotted in the inset of \textbf{c}.
			On-site potential disorder $V_j$ follows a uniform random distribution within the range $[-V, V]$. The red solid curve represents the mean numerical results, and the error bars represent the standard error of the experimental (numerical) results with $30$~($100$) configurations of disorder.
			\textbf{b}, Experimental data of $\delta x$ versus $V$ during a double-loop pumping trajectory $\mathcal{C}_{\text{dl}}$ as shown in the inset of \textbf{e}.
			\textbf{c}, Numerical results of the charge pumped per cycle
			$\Delta Q$ versus $V$ during the outer-loop pumping trajectory $\mathcal{C}_{\text{out}}$.
			\textbf{d}, Numerical results of $\Delta Q$ versus $V$ during the inner-loop pumping trajectory $\mathcal{C}_{\text{in}}$.
			\textbf{e}, $\Delta Q$ for double-loop pumping   $\mathrm{C}_{\text{dl}}$, which is obtained by summing the results of outer- and inner-loop pumping.
			\textbf{f}, Bulk energy band for double-loop pumping under open boundary conditions. Darker colors imply higher state density.
			\textbf{g}, Experimental data of the average populations of all qubits  during the adiabatic cyclic evolution for double-loop pumping over 30 independent disorder configurations. The period of both the outer- and the inner-loop pumping is set as $500$~ns.}
		\label{fig:fig3}
	\end{figure*}

	In our experiments, we engineer the continuous RM model in the clean limit and implement topological pumping by periodically modulating $\delta$ and $\Delta$ that sketch a closed trajectory
	in the $\delta$--$\Delta$ space within a period $T$. The initial state is prepared as a single-excitation state, having an overlap of over $0.99$ with the exact Wannier state~\cite{supp_cite}, by exciting one qubit closest to the middle, i.e., $Q_{19}$. During the pumping procedure, we measure the population of each qubit $P_{|1\rangle}\equiv\langle\hat{n}_j\rangle$, with which the CoM can be calculated as $ \bar{x}\equiv\sum_j{j}\langle\hat{n}_j\rangle$.
	The experimental results of the shift of the CoM after four pumping cycles  are shown in
	Fig.~\ref{fig:fig2} for four distinct pumping trajectories $\mathcal{C}_1$--$~\mathcal{C}_4$ (Fig.~\ref{fig:fig2}c), respectively.
	The period is carefully chosen as $T=650$~ns, when the mean $\delta x$  achieves its maximum $1.95(6)$ {(Fig.~\ref{fig:fig2}b)}. Here, the slight oscillation of $\delta x$ for $T>650$~ns originates from the difference between the single-excitation state and the exact Wannier state. 
	Quantized charge pumping is observed for topologically nontrivial pumping trajectories $\mathcal{C}_1$ and $~\!\mathcal{C}_4$  around the gapless point $(\Delta, \delta)=(0, 0)$, corresponding to the Chern numbers $\mp1$, respectively.
	Moreover, topologically trivial pumping is probed for $\mathcal{C}_2$ and $\mathcal{C}_3$ with zero Chern number. The corresponding energy bands under open boundary conditions  are shown in Fig.~\ref{fig:fig2}d for $\mathcal{C}_1$--$~\mathcal{C}_4$, respectively, which could be measured by a dynamical spectroscopic technique~\cite{shi2022observing}. The deviation for $t>3T$, between the experimental and numerical results in Fig.~\ref{fig:fig2}a, are due to dephasing~\cite{supp_cite}. Adiabatic time evolutions for a pumped excitation during pumping trajectories $\mathcal{C}_1$--$\mathcal{C}_4$ are shown in Fig.~\ref{fig:fig2}e.  
	In addition, we experimentally monitor the  double-excitation pumps for different trajectories~\cite{supp_cite}, which are shown in Fig.~\ref{fig:figdouble}. The experimental results are similar to the single-excitation cases, as the system is in the hard-core limit~\cite{yan_strongly_2019}. Since the pumps of  excitations initially prepared at odd and even sites have opposite winding numbers~\cite{ke_multiparticle_2017}, no quantized pumping is observed for the topologically nontrivial pumping trajectory $\mathcal{C}_4$, when the parity of the initial excitation sites is different (Fig.~\ref{fig:figdouble}d).
	\\	
	\\
	\noindent \textbf{Pumping in the presence of on-site disorder}
	 \\
	Next, we investigate the effects of on-site potential disorder on topological pumping. Figure~\ref{fig:fig3}a shows the displacement of the CoM for a forward pump, with respect to the pumping trajectory $\mathcal{C}_{\text{out}}$ (inset of Fig.~\ref{fig:fig3}c), versus the on-site disorder strength $V/2\pi$.
	Here, the on-site potential disorder $V_j$ on each qubit satisfies a uniform random distribution in the range $[-V, V]$. The experimental results demonstrate that
	quantized pumping persists for $V/\Delta_0\lesssim 1$, 
	but degrades as the displacement of the CoM per pumping cycle $\delta x$ decays to zero for $V\gtrsim 3\Delta_0$.
	In addition, we numerically calculate the pumping amounts of charge over one cycle, i.e.,
	\begin{equation}
		\Delta Q=d\!\int_0^T\!\!\!\mathrm{d}t\;\bra{\psi(t)}\hat{\mathcal{J}}(t)\ket{\psi(t)},
	\end{equation}
	to characterize the interplay between topology and disorder~\cite{wauters_localization_2019, wu_quantized_2022}
	(Fig.~\ref{fig:fig3}c--e). Here, the average current density can
	be expressed as
	\begin{equation}\label{eqJ}
		\hat{\mathcal{J}}(t)=\mathrm{i}\sum_{j=1}^N[(J+(-1)^{j-1}\delta)\hat{a}_{j+1}^{\dagger}\hat{a}_{j}+\mathrm{H.c.}]/N,
	\end{equation}
	and $\ket{\psi(t)}$ is the time evolved state initially with a half-filling ground state of the system, and $\mathrm{i}=\sqrt{-1}$. As shown in Fig.~\ref{fig:fig3}c, $\Delta Q$ versus $V$ has a similar behavior as the experimental results of $\delta x$. The slight reduction of $\Delta Q$ when $V/\Delta_0\lesssim 1$ results from the use of a single-excitation initial state instead of an exact Wannier state. The breakdown of quantized pumping can be understood due to the closing band gap, leading to the Landau-Zener transition~\cite{niu_quantised_1984, ivakhnenko_nonadiabatic_2023}. The gap closes when $V\approx \Delta_0$~\cite{supp_cite}, which conforms to the experimental observations of $\delta x$. Thus, the breakdown may be due to  localization of single-particle Floquet states instead of that of instantaneous eigenstates, for which localization occurs for any non-zero disorder strength~\cite{wauters_localization_2019}.
	
	\begin{figure*}[t!]
		\centering
		\includegraphics[width=0.97\linewidth]{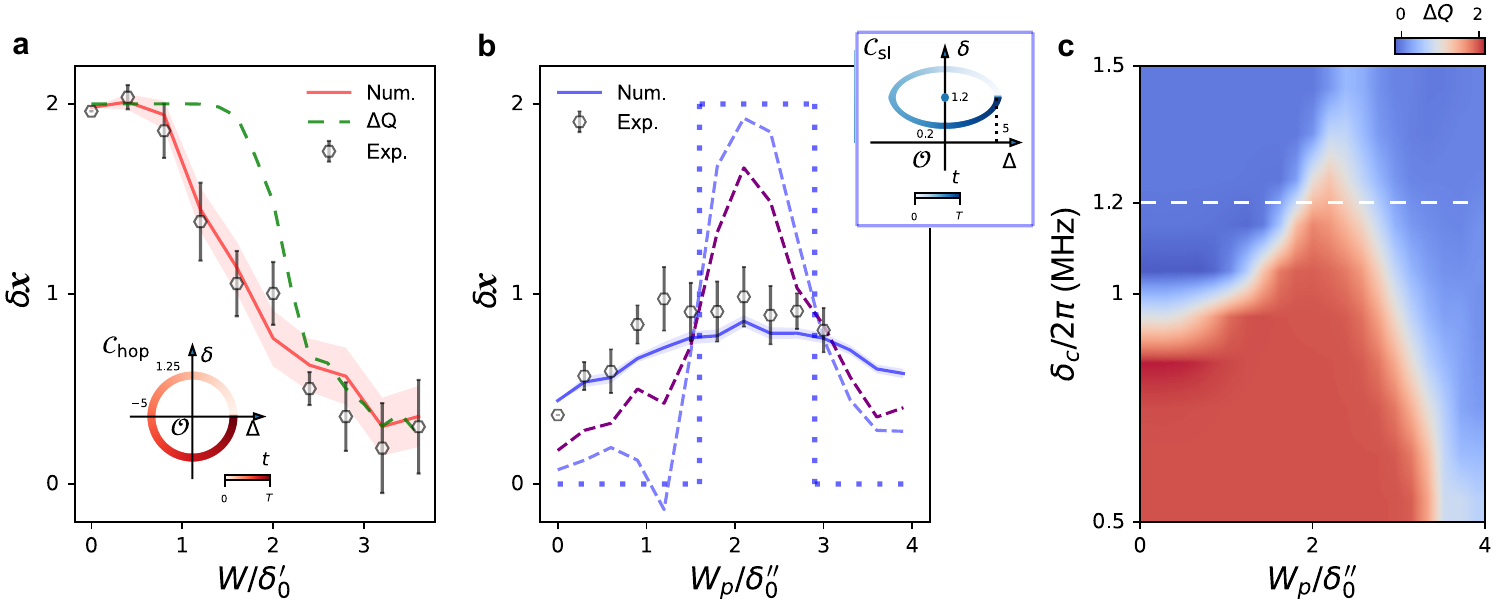}
		\caption{\textbf{Pumping with hopping disorder}. \textbf{a}, Experimental data for the displacement of the CoM $\delta x$ for one pumping period versus the random hopping disorder strength $W$ during the pumping trajectory $\mathcal{C}_{\text{hop}}$ as shown in the inset.
			The red solid and green dashed curves plot the numerical results of $\delta x$ and $\Delta Q$, respectively. \textbf{b}, Experimental data of $\delta x$ against the quasi-periodic disorder strength $W_p$ during the pumping trajectory $\mathcal{C}_{\text{sl}}$:
			$(\Delta, \delta)=(\Delta_0'\cos 2\pi t/T, \delta_{c}+\delta''_0\sin 2\pi t/T)$, with $\Delta'_0/2\pi=5~\mathrm{MHz}$, $\delta_c/2\pi=1.2~\mathrm{MHz}$, $\delta''_0/2\pi=1~\mathrm{MHz}$, $T=1.4$~$\mu$s, and
			$J/2\pi=1.8~\mathrm{MHz}$. The solid blue curve denotes the numerical simulation and the dotted curve shows the topological index calculated in the thermodynamic limit~\cite{mondragon_shem_disorder_2014}. The purple (light blue) dashed curve denotes the numerical results using similar experimental parameters but with a longer period $20$~$\mu$s ($80$~$\mu$s) in a larger system with $200$ ($800$) qubits. Experimental data are averaged over 30 disorder configurations, while the numerical simulation is calculated for 1,000 disorder configurations. \textbf{c}, Charge pumped per cycle $\Delta Q$ versus $W_p$ and $\delta_c$. The white dashed horizontal line shows the TAI-like topological transition of pumping during $\mathcal{C}_{\text{sl}}$.}
		\label{fig:fig4}
	\end{figure*}

	In addition, we demonstrate a pumping procedure following a double-loop pumping trajectory, $\mathcal{C}_{\text{dl}}$, to study topological pumping that is induced by random on-site disorder. As plotted in the inset of Fig.~\ref{fig:fig3}e, this closed pumping trajectory is composed of an outer loop $\mathcal{C}_{\text{out}}$ (inset of Fig.~\ref{fig:fig3}c) and an inner loop $\mathcal{C}_{\text{in}}$ (inset of Fig.~\ref{fig:fig3}d).
	Since along $\mathcal{C}_{\text{out}}$ and $\mathcal{C}_{\text{in}}$, the system evolves into a parameter plane with opposite directions, there is no net pumped charge with zero total Chern number as $\nu_{\text{out}}+\nu_{\text{in}}=0$.
	However, as the on-site disorder strength $V$ increases, the gapless regime appears around the origin
	$\mathcal{O}$ along the $\Delta$-axis.
	When $0.5\lesssim{V/\Delta_0}\lesssim2$,
	the inner loop cannot encircle the whole gapless regime and no topological pumping phenomenon occurs, while the outer loop remains nontrivial with $\nu_{\text{out}}=+1$.
	Thus, with a moderate disorder strength, we observe nontrivial pumping induced by the on-site disorder with $\delta x\neq{0}$  (Fig.~\ref{fig:fig3}b). 
	However, a quantized disorder-induced pump can hardly be realized, since quantized transport requires trajectory parameters to be finely tuned to combine the effects of the trivial inner and outer trajectories ~\cite{nakajima_competition_2021}.
	As the disorder strength increases further to $V/\Delta_0\gtrsim2.5$, pumping becomes trivial, since no topological pumping exists for the outer loop. 
	The increase of $\delta x$ in the region $0\lesssim V/\Delta_0\lesssim 0.7$ is also due to the discrepancy between the single-excitation and  Wannier initial states.
	\\
	\\
	\noindent \textbf{Pumping with hopping disorder}
	\\
	Next, we experimentally investigate topological pumping in the presence of hopping disorder.
	We choose a trivial pumping trajectory $\mathcal{C}_{\text{hop}}$:
	$\displaystyle (\Delta,\delta)=(\Delta'_0\cos 2\pi t/T, \delta'_0\sin 2\pi t/T)$,
	with $\Delta'_0/2\pi=5$~MHz, $\delta'_0/2\pi=1.25$~MHz, $J/2\pi=1$~MHz, and $T=1.3$~${\mu}$s (inset of Fig.~\ref{fig:fig4}a).
	First, we consider uniform random hopping disorder within the range $[-W, W]$.
	The experimental results, shown in Fig.~\ref{fig:fig4}a, are similar to the ones with on-site potential disorder as shown in Fig.~\ref{fig:fig3}a, where the increase of the disorder strength leads to the decrease of $\delta x$.
	However, the decay of $\delta x$ obeys a distinct law from the on-site potential disorder case \cite{soukoulis_off-diagonal_1981}, when the 1D system tends to the localization phase.
	A non-adiabatic evolution could cause the breakdown of quantized pumping with a smaller disorder strength, which is verified by comparing the charge pumped with a longer period with the transition point at $W/\delta_0\approx1$~\cite{supp_cite}.
	
	Recently, it has been suggested that quasi-periodic hopping disorder would lead to exotic topological phenomena \cite{tang_topological_2022}. Moreover, as the gap would reopen, applying quasi-periodic hopping disorder may intrinsically induce topological pumping, which can hardly be realized by introducing random hopping disorder~\cite{supp_cite, wu_quantized_2022}. Here, we consider a topologically trivial single-loop  pumping trajectory with its center being biased away from the gapless point $\mathcal{O}$  (origin of $\Delta$--$\delta$ plane), i.e., $\mathcal{C}_{\text{sl}}$:
	$\displaystyle (\Delta, \delta)=(\Delta'_0\cos 2\pi t/T, \delta_{c}+\delta''_0\sin 2\pi t/T)$ with $\Delta'_0/2\pi=5$~{MHz}, $\delta_c/2\pi=1.2~\mathrm{MHz}$, $\delta''_0/2\pi=1$~{MHz}, $T=1.4$~$\mu$s,
	and $J/2\pi=1.8$~{MHz} (inset of Fig.~\ref{fig:fig4}b).
	Quasi-periodic hopping disorder,
	$\displaystyle W_{j}=W_p\cos(2\pi\alpha j+{\color{blue}\beta})$, is introduced on each even qubit, with $\alpha=(\sqrt{5}-1)/2$ being irrational and $\beta\in[-\pi, \pi)$ being an arbitrary random phase offset.
	As the disorder strength $W_p$ increases, the gapless point would appear inside the pumping loop \cite{wu_quantized_2022}, and nontrivial pumping could be observed (see the theoretical predictions in Fig.~\ref{fig:fig4}b).
	Though under insufficient adiabaticity, we demonstrate the observation of
		signatures consistent with topological pumping induced by quasi-periodic hopping
		disorder, which leads to nonzero $\delta x$ in the clean limit. Theoretically, with an extremely long evolution period, e.g., $20$~$\mu$s  and $80$~$\mu$s,
	as shown in Fig.~\ref{fig:fig4}b, non-adiabatic effects can be suppressed.
	Moreover, this nontrivial pumping phenomenon 
	could also be viewed as a dynamical version of TAIs~\cite{meier_observation_2018,stutzer_tai_2018}, as the numerical results of $\Delta Q$ in Fig.~\ref{fig:fig4}c  indicates the existence of TAI-like topological transitions.
	\\
	\\
	\noindent \textbf{Outlook}
	\\
	We experimentally investigated the competition and interplay between topology and disorder in topological pumping on a 41-qubit superconducting processor.  Furthermore, we demonstrated disorder-induced topological pumping which was induced by either on-site random disorder or quasi-periodic hopping disorder. 
	In addition, we experimentally studied the robustness and the breakdown of a Thouless pump as the disorder strength increases.
	Note that these experimental results were obtained by extending the multi-qubit Floquet engineering technique to the adiabatic evolution regime, which would be helpful in exploring various topological phenomena induced by disorder.
	\\
	\\
	\noindent \textbf{Methods}
	\\
	\noindent \textbf{Floquet engineering for adiabatic systems}
	\\
	In our experiments, we employ an extended Floquet engineering technique with the high-frequency expansion~\cite{oka_floquet_2019} to realize the RM model, which is an effective approach to modulate hopping strengths between qubits. Since the simultaneous changes of on-site potentials and hopping strengths are inherently necessary, we extended Floquet engineering for adiabatic systems, by carefully introducing two restrictions: the adiabatic condition and the Nyquist condition. Specifically, we manipulate the Z pulse to tune the $j$-th qubit frequency according to
	\begin{equation}\label{eq:freq_wave}
		\omega_j(t)=\bar{\omega}+\Delta_j(t)+A_j(t)\sin(\mu t+\varphi_0),
	\end{equation}
	where $\bar{\omega}$ is the average frequency, $A_j$, $\mu$, and $\varphi_0$ denote modulation amplitude, frequency, and phase, respectively, and $\Delta_j$ is the $j$-th on-site potential. Experimentally, we set $\bar{\omega}/2\pi=4.8$~GHz, and $\mu/2\pi=80$~MHz for all qubits,
	and a schematic of the qubit frequency is plotted in Fig.~\ref{fig:wavediagram}a. To realize the high-frequency expansion, the modulation frequency should be higher than the simulated frequency regime for fulfilling the adiabatic condition, and the effective Hamiltonian contains a series of frequency bands. The Nyquist condition requires that the variation range of the difference between two neighbor on-site potentials should be lower than half the modulation frequency $\mu/2$. This can avoid any overlap between different frequency bands, resulting in an effective simulation of the target time-evolved Hamiltonian under the rotating wave approximation.
	
	By introducing the superconducting quantum interference device (SQUID) into the transmon qubit, the qubit is frequency-tunable, and the relationship between the qubit frequency $\omega$ and the flux $\Phi_e$, entering the loop of SQUID~\cite{koch_charge-insensitive_2007}, is
	\begin{equation}\label{eq:zpa2w}
		\omega=\sqrt{8E_{JJ}E_C|\cos(\pi{\Phi_e}/{\Phi_0})|}-E_C,
	\end{equation}
	where $E_{JJ}$ denotes the Josephson energy when $\Phi_e=0$, $E_C$ is the charging energy, and $\Phi_0$ is the flux quantum. For weak magnetic fields, $\Phi_e$ is linearly related to the experimental Z pulse amplitude (Zpa) $V_z$, i.e., $\pi\Phi_e/\Phi_0=kV_z+b$.
	These parameters can be extracted by the single-qubit spectroscopy measurement experiments. However, the parameters obtained in this way would be inaccurate due to the unavoidable crosstalk after tuning all qubits to their idle points. Thus, we apply the multi-qubit spectroscopy measurements in the range near the working points or the average frequency $\bar{\omega}\sim4.8$~GHz, see Fig.~\ref{fig:wavediagram}b. Then, we fit the relationship in Eq.~\eqref{eq:zpa2w} using this small segment of the spectroscopy data, which exhibits a linear correlation. Although under-fitting seems to occur, we could achieve the desired results by fixing the known parameters insensitive to the crosstalk, such as $E_C$ and the sweet points of qubits. The inset of Fig.~\ref{fig:wavediagram}b shows the optimized mapping from Zpa to the qubit frequency, which differs from single-qubit fitting result.
	
	Combining Eqs.~\eqref{eq:freq_wave} and \eqref{eq:zpa2w}, we can obtain the Z pulse waveform, applied on the $j$-th qubit $V_j^z$, as 
		$V_z^j=\frac{1}{k_j}\arccos\left[\pm\frac{(\bar{\omega}+\Delta_j(t)+A_j(t)\sin(\mu t+\varphi_0)+E_C^j)^2}{8E_{JJ}^jE_C^j}\right]-\frac{b_j}{k_j}$.
	Note that $A_j(t)$ is dependent of the modulation amplitude of the nearest qubits $A_{j-1}(t)$ and $A_{j+1}(t)$. In practical operations, we establish a reference amplitude, which is a smooth function, for a specific qubit $Q_k$. For convenience,  we simply set $A_k(t)\equiv 0$, and then, we perform the iterative calculation of $Q_k$ to obtain $A_{k+1}(t), A_{k+2}(t), \cdots$ and $A_{k-1}(t), A_{k-2}(t), \cdots$.
	
	Using the method as introduced above, we can engineer a time-dependent Hamiltonian with the simultaneous adjustment of the on-site potentials and the hopping strengths on our superconducting processor with only frequency-tuning capabilities. Numerically, we calculate pumping for the trajectory, $\mathcal{C}_{4}$, by evolving the exact RM model as shown in Fig.~\ref{fig:calculatefloquetvsexact}a and the same Hamiltonian, but constructed through Floquet engineering, as shown in Fig.~\ref{fig:calculatefloquetvsexact}b, respectively. The CoM extracted from these two methods coincide very well, see Fig.~\ref{fig:calculatefloquetvsexact}c.\\
	
\noindent \textbf{Experimental setup}
	\\
	Our superconducting quantum processor consists $43$ transmon qubits arranged in a $1$D array, labeled as $Q_1$, $\cdots$, $Q_{43}$, and we used $Q_3$, $\cdots$, $Q_{43}$ (relabeled as $Q_1$, $\cdots$, $Q_{41}$) for the experiments.The qubits are capacitively coupled to their
	nearby qubits with a mean hopping strength $\overline{g}/2\pi\simeq7.2$~MHz, which suggests that the adjustable range of the effective hopping strengths is from $-2.9$~MHz to $7.2$~MHz. Since the average anharmonicity is $\overline{U}/2\pi\simeq-208$~MHz, with a ratio $|\overline{U}/\overline{g}|\simeq29\gg1$, our processor can  be regarded as a hard-core bosonic system \cite{yan_strongly_2019}.
	The mean energy relaxation time is $21.0$~$\mu s$, and the sweet points of qubits are designed to be staggered  for the convenience of arranging energy levels, with a mean value of $5.014$~GHz.  
	
	With all 41 superconducting qubits initialized at their idle points, we prepared the localized initial state using an X gate, as an approximation to the Wannier state. By using the derivative removal by adiabatic gate (DRAG) theory \cite{motzoi_simple_2009}, the X gate pulse is optimized to minimize the leakage to higher energy levels, achieving an average gate fidelity of 99.2\%. Then, the parametric flux modulations are applied on all qubits to engineer the Rice-Mele Hamiltonian, for different pumping experiments. The schematic diagram of the pulse sequence, for the double-excitation experiments as an example, is shown in Fig.~\ref{fig:methodsfig3}. After turning off the parametric driving, the qubits are tuned back to their idle points for readout. The states of all qubits can be read out simultaneously through the transmission lines coupled to readout resonators. All qubit probabilities are corrected to eliminate the measurement errors.
	\\
	
\noindent \textbf{Acknowledgments}
We thank Yun-Long Su for the helpful discussions and the support from the Synergetic Extreme Condition User Facility (SECUF) in Huairou District, Beijing. 
Devices were made at the Nanofabrication Facilities at Institute of Physics, CAS in Beijing.
	\\
	
	
	\noindent\textbf{Author contributions} H.F. and K.X. supervised the project; Y.-R.Z. proposed the idea; Y.L. performed the experiment with the assistance of Y.-H.S. and K.X.; Z.X. fabricated the device with the help of G.-H.L., Z.-Y.M., and D.Z.; Y.L., C.L. and S.-Y.Z. performed the numerical simulations and discussed with Y.-R.Z., Y.-H.S., Tao L., Tong.L., Y.-Y.W., and K.X.;
	H.L., T.-M.L., C.-L.D., Tong L., J.-C.Z., G.-H.L., Z.-Y.M., W.-G.M., H.-T.L., Z.-H.L, C.-T.C., K.H., S.P.Z., and Y.T. helped the experimental setup supervised by K.X.; X.S. provided the Josephson parametric amplifiers; Y.-R.Z., F.N., and H.F. gave theoretical explanations; Y.L., Y.-H.S., Y.-R.Z, K.X., and H.F. co-wrote the manuscript, and all authors contributed to the discussions of the results and development of the manuscript.\\

	\noindent\textbf{Competing interests:} The authors declare no competing interests.\\
	
	\noindent\textbf{Data and materials availability:} All data needed to evaluate the conclusions in the paper are present in the paper and/or the Supplementary Materials.
	
	\newpage
	
	\bibliography{main}
	~\\

	\setcounter{figure}{0}
	\renewcommand{\figurename}{\textbf{Extended Data Fig.}}

	\begin{figure*}[htb]
		\centering
		\includegraphics[width=0.97\linewidth]{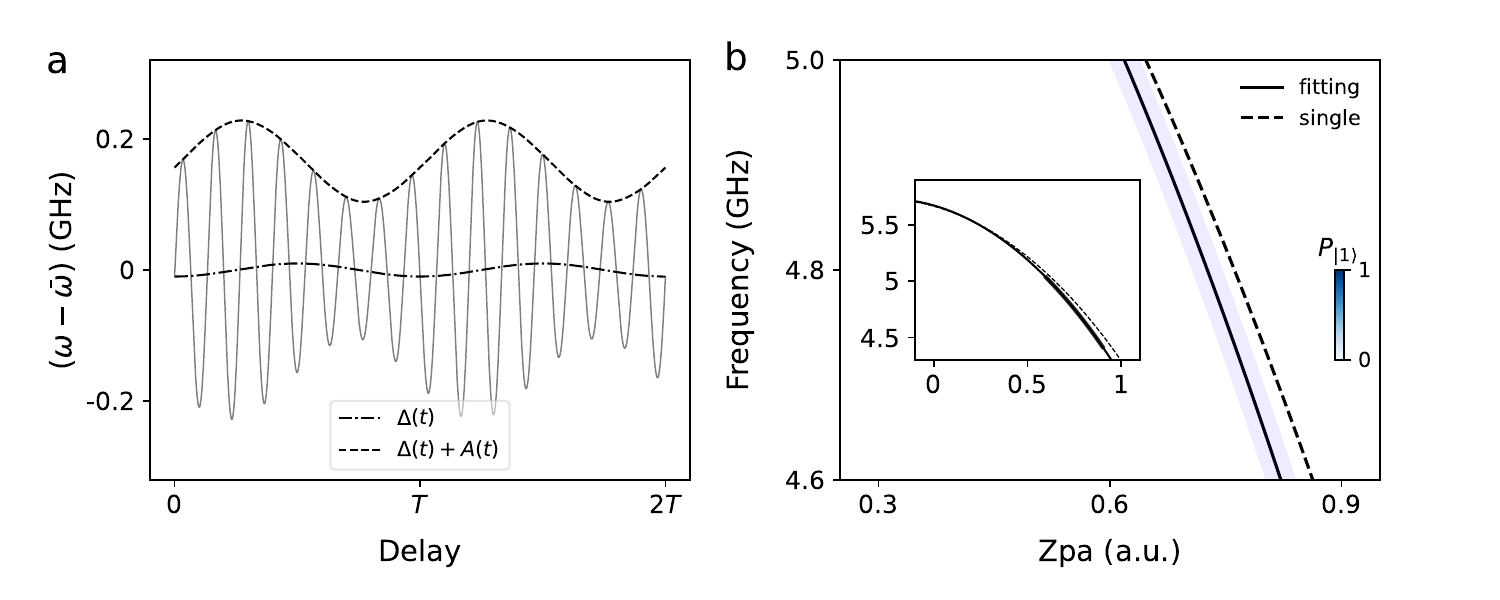}
		\caption{\textbf{Pulse sequence}.
			\textbf{a}, Schematic of the qubit frequency. The frequency is manipulated as amplitude-modulated sinusoidal wave with a moving center.
			\textbf{b}, Multi-qubit spectroscopy measurement. The dashed line represents the mapping from Z pulse amplitude (Zpa) to qubit frequency obtained by single-qubit spectroscopy. The solid line shows fitting result from the multi-qubit spectroscopy segment. Due to crosstalk, the dashed line deviates from the actual mapping when tuning all qubits to their work points, indicated by curves in the inset.}
		\label{fig:wavediagram}
	\end{figure*}
	
	\begin{figure*}[t]
		\centering
		\includegraphics[width=0.97\linewidth]{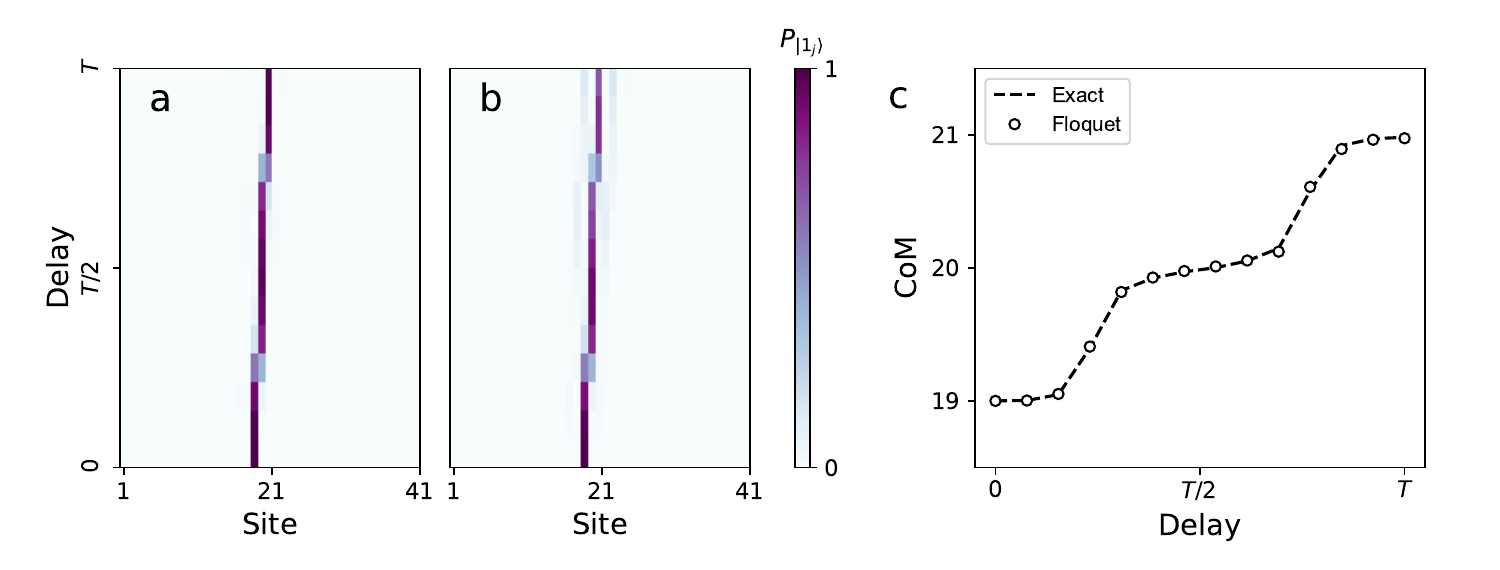}
		\caption{\textbf{Numerical calculation of pumping for $\mathcal{C}_{4}$}.
				\textbf{a}, Numerical results of the time evolution of the initial single-excitation state obtained by evolving the exact Rice-Mele model Hamiltonian over one cycle.
				\textbf{b}, Numerical results of the time evolution of the initial single-excitation state obtained by evolving the Rice-Mele model Hamiltonian constructed through Floquet engineering over one cycle. \textbf{c}, Numerical results of CoM extracted from \textbf{a} and \textbf{b}. The initial state is chosen as a single-excitation state at the $19$-th site, and the modulation frequency $\mu$ is set as
				$\mu/2\pi=80$~MHz.}
		\label{fig:calculatefloquetvsexact}
	\end{figure*}

	\clearpage
	\begin{figure*}
		\centering
		\includegraphics[width=0.85\linewidth]{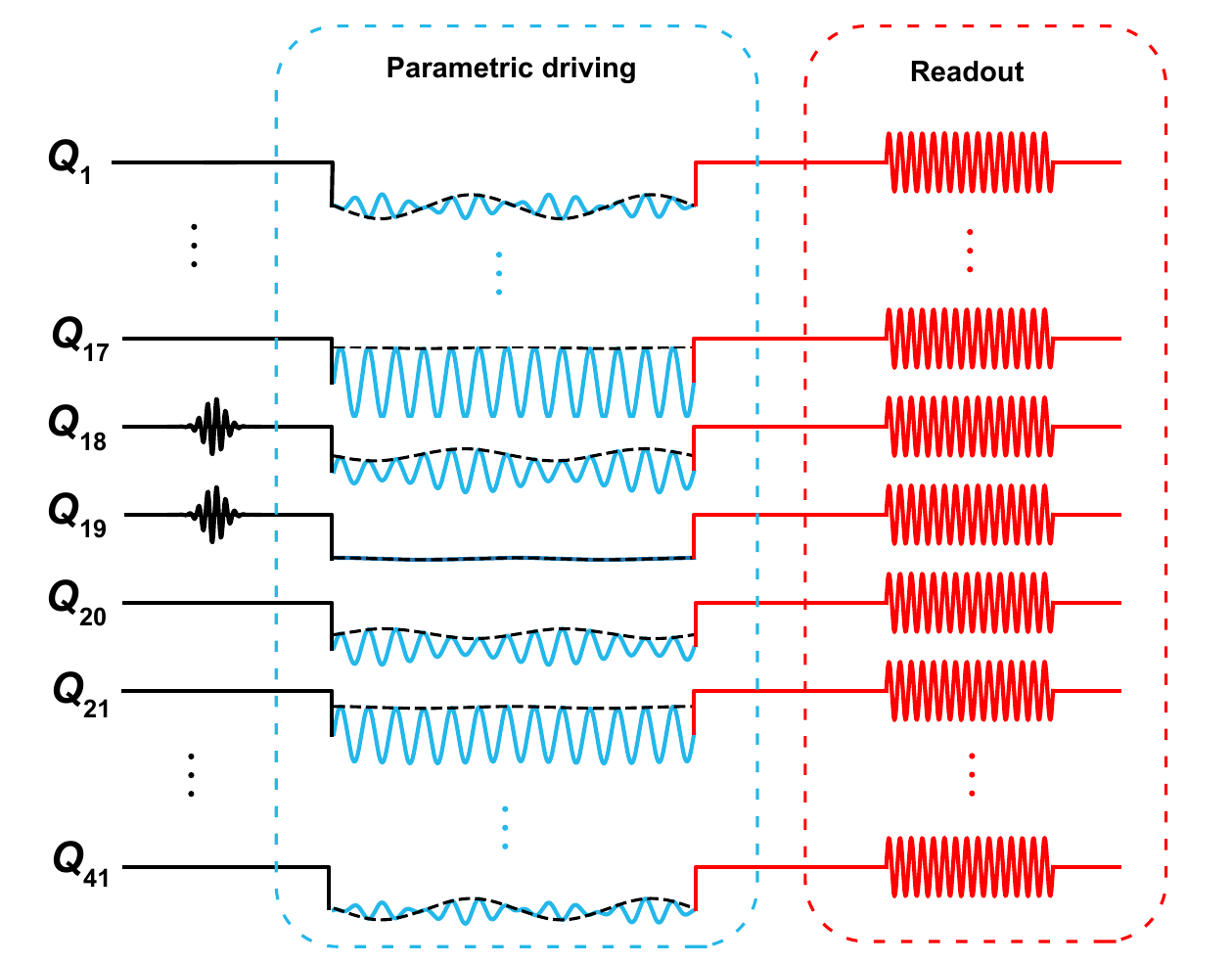}
		\caption{\textbf{Pulse sequence for topological pumping with double excitations.} All qubits are initialized at their idle points. Then, $Q_{18}$ and $Q_{19}$ are excited using the Gaussian-like derivative removal by adiabatic gate (DRAG) pulses. Next, all qubits are driven by performing parametric modulations through their Z-control lines. After a delay time, the parametric driving is turned off, and all qubits are tuned back to their idle points for readouts.}
		\label{fig:methodsfig3}
	\end{figure*}

\end{document}


\title{Supplementary Materials for:\\ ``Interplay between disorder and topology in Thouless pumping on a superconducting quantum processor"}
	
	\author{Yu Liu}
	\thanks{These authors contributed equally to this work.}
	\affiliation{Institute of Physics, Chinese Academy of Sciences, Beijing 100190, China}
	\affiliation{School of Physical Sciences, University of Chinese Academy of Sciences, Beijing 100049, China}
	
	\author{Yu-Ran Zhang}
	\thanks{These authors contributed equally to this work.}
	\affiliation{School of Physics and Optoelectronics, South China University of Technology, Guangzhou 510640, China}
	
	\author{Yun-Hao Shi}
	\affiliation{Institute of Physics, Chinese Academy of Sciences, Beijing 100190, China}
	\affiliation{School of Physical Sciences, University of Chinese Academy of Sciences, Beijing 100049, China}
	\affiliation{Beijing Academy of Quantum Information Sciences, Beijing 100193, China}
	
	\author{Tao Liu}
	\affiliation{School of Physics and Optoelectronics, South China University of Technology, Guangzhou 510640, China}
	
	\author{Congwei Lu}
	\affiliation{Department of Physics, Applied Optics Beijing Normal University, Beijing 100875, China}
	
	\author{Yong-Yi Wang}
	\affiliation{Institute of Physics, Chinese Academy of Sciences, Beijing 100190, China}
	\affiliation{School of Physical Sciences, University of Chinese Academy of Sciences, Beijing 100049, China}
	
	\author{Hao Li}
	\affiliation{Institute of Physics, Chinese Academy of Sciences, Beijing 100190, China}
	
	\author{Tian-Ming Li}
	\affiliation{Institute of Physics, Chinese Academy of Sciences, Beijing 100190, China}
	\affiliation{School of Physical Sciences, University of Chinese Academy of Sciences, Beijing 100049, China}
	
	\author{Cheng-Lin Deng}
	\affiliation{Institute of Physics, Chinese Academy of Sciences, Beijing 100190, China}
	\affiliation{School of Physical Sciences, University of Chinese Academy of Sciences, Beijing 100049, China}
	
	\author{Si-Yun Zhou}
	\affiliation{Institute of Physics, Chinese Academy of Sciences, Beijing 100190, China}
	\affiliation{School of Physical Sciences, University of Chinese Academy of Sciences, Beijing 100049, China}
	
	\author{Tong Liu}
	\affiliation{Institute of Physics, Chinese Academy of Sciences, Beijing 100190, China}
	
	\author{Jia-Chi Zhang}
	\affiliation{Institute of Physics, Chinese Academy of Sciences, Beijing 100190, China}
	\affiliation{School of Physical Sciences, University of Chinese Academy of Sciences, Beijing 100049, China}
	
	\author{Gui-Han Liang}
	\affiliation{Institute of Physics, Chinese Academy of Sciences, Beijing 100190, China}
	\affiliation{School of Physical Sciences, University of Chinese Academy of Sciences, Beijing 100049, China}
	
	\author{Zheng-Yang Mei}
	\affiliation{Institute of Physics, Chinese Academy of Sciences, Beijing 100190, China}
	\affiliation{School of Physical Sciences, University of Chinese Academy of Sciences, Beijing 100049, China}
	
	\author{Wei-Guo Ma}
	\affiliation{Institute of Physics, Chinese Academy of Sciences, Beijing 100190, China}
	\affiliation{School of Physical Sciences, University of Chinese Academy of Sciences, Beijing 100049, China}
	
	\author{Hao-Tian Liu}
	\affiliation{Institute of Physics, Chinese Academy of Sciences, Beijing 100190, China}
	\affiliation{School of Physical Sciences, University of Chinese Academy of Sciences, Beijing 100049, China}
	
	\author{Zheng-He Liu}
	\affiliation{Institute of Physics, Chinese Academy of Sciences, Beijing 100190, China}
	\affiliation{School of Physical Sciences, University of Chinese Academy of Sciences, Beijing 100049, China}
	
	\author{Chi-Tong Chen}
	\affiliation{Institute of Physics, Chinese Academy of Sciences, Beijing 100190, China}
	\affiliation{School of Physical Sciences, University of Chinese Academy of Sciences, Beijing 100049, China}
	
	\author{Kaixuan Huang}
	\affiliation{Beijing Academy of Quantum Information Sciences, Beijing 100193, China}
	
	\author{Xiaohui Song}
	\affiliation{Institute of Physics, Chinese Academy of Sciences, Beijing 100190, China}
	
	\author{S. P. Zhao}
	\affiliation{Institute of Physics, Chinese Academy of Sciences, Beijing 100190, China}
	\affiliation{School of Physical Sciences, University of Chinese Academy of Sciences, Beijing 100049, China}
	\affiliation{Songshan Lake  Materials Laboratory, Dongguan, Guangdong 523808, China}
	
	\author{Ye Tian}
	\affiliation{Institute of Physics, Chinese Academy of Sciences, Beijing 100190, China}
	
	\author{Zhongcheng Xiang}
	\email{zcxiang@iphy.ac.cn}
	\affiliation{Institute of Physics, Chinese Academy of Sciences, Beijing 100190, China}
	\affiliation{School of Physical Sciences, University of Chinese Academy of Sciences, Beijing 100049, China}
	\affiliation{Hefei National Laboratory, Hefei 230088, China}
	
	\author{Dongning Zheng}
	\affiliation{Institute of Physics, Chinese Academy of Sciences, Beijing 100190, China}
	\affiliation{School of Physical Sciences, University of Chinese Academy of Sciences, Beijing 100049, China}
	\affiliation{Songshan Lake Materials Laboratory, Dongguan, Guangdong 523808, China}
	\affiliation{CAS Center for Excellence in Topological Quantum Computation, UCAS, Beijing 100049, China}
	\affiliation{Hefei National Laboratory, Hefei 230088, China}
	
	\author{Franco Nori}
	\affiliation{Theoretical Quantum Physics Laboratory, Cluster for Pioneering Research, RIKEN, Wako-shi, Saitama 351-0198, Japan}
	\affiliation{Center for Quantum Computing, RIKEN, Wako-shi, Saitama 351-0198, Japan}
	\affiliation{Physics Department, University of Michigan, Ann Arbor, Michigan 48109-1040, USA}
	
	\author{Kai Xu}
	\email{kaixu@iphy.ac.cn}
	\affiliation{Institute of Physics, Chinese Academy of Sciences, Beijing 100190, China}
	\affiliation{School of Physical Sciences, University of Chinese Academy of Sciences, Beijing 100049, China}
	\affiliation{Beijing Academy of Quantum Information Sciences, Beijing 100193, China}
	\affiliation{Songshan Lake Materials Laboratory, Dongguan, Guangdong 523808, China}
	\affiliation{CAS Center for Excellence in Topological Quantum Computation, UCAS, Beijing 100049, China}
	\affiliation{Hefei National Laboratory, Hefei 230088, China}
	
	\author{Heng Fan}
	\email{hfan@iphy.ac.cn}
	\affiliation{Institute of Physics, Chinese Academy of Sciences, Beijing 100190, China}
	\affiliation{School of Physical Sciences, University of Chinese Academy of Sciences, Beijing 100049, China}
	\affiliation{Beijing Academy of Quantum Information Sciences, Beijing 100193, China}
	\affiliation{Songshan Lake Materials Laboratory, Dongguan, Guangdong 523808, China}
	\affiliation{CAS Center for Excellence in Topological Quantum Computation, UCAS, Beijing 100049, China}
	\affiliation{Hefei National Laboratory, Hefei 230088, China}
	\maketitle
	\beginsupplement
	\tableofcontents
	\newpage

	\section{Experimental setup}
	Our experiments are performed on \emph{Chuang-tzu}, a 1D superconducting  processor, containing 43 transmon qubits,
	which is the same processor used in~\cite{shi2022observing}. The qubits are designed to be frequency-tunable with a mean sweet point of about \SI{5.8}{\GHz},
	and each qubit is capacitively coupled to its nearby qubits with a mean coupling strength of about \SI{7.2}{\MHz}. We use 41 qubits, i.e.,
	$Q_1$--$Q_{41}$, in our experiments, and all relevant information about qubit characteristics is listed in Table~\ref{tab:paras}.
	All qubits are initialized at their idle frequencies distributing over the range from \SI{4.4}{\GHz} to \SI{5.6}{\GHz},
	which are carefully arranged to minimize the unexpected interaction and crosstalk errors. The anharmonicity, $\alpha_j/2\pi$,
	of qubits is about \SI{0.2}{\GHz}, and the qubit working frequency is adjusted to be \SI{4.8}{\GHz}.
	We tune all qubit frequencies and coupling strengths to the target points by applying the automatic calibration scheme as mentioned in Ref.~\cite{shi2022observing}.
	
	\begin{table*}[b]
		\begin{ruledtabular}
			\begin{tabular}{lcccc}
				Parameter & Mean & Median & Stdev. & Units\\[.05cm] \hline
				\\[-.2cm]
				Qubit maximum frequency $\omega_{\textrm{m}}/2\pi$& $5.83$ & $5.768$ & $0.289$ & GHz \\[.09cm]
				Qubit idle frequency $\omega_{\textrm{i}}/2\pi$ & $5.014$ & $4.777$ & $0.428$ & GHz \\[.09cm]
				Qubit anharmonicity $\alpha_j/2\pi$ & $-0.208$ & $-0.193$ & $0.02$ & GHz\\[.09cm]
				Readout frequency $\omega_{\textrm{r}}/2\pi$ & $6.68$ & $6.684$ & $0.052$ & GHz\\[.09cm]
				Mean energy relaxation time $\overline{T}_1$ & $21.0$ & $20.9$ & $6.0$ & \us\\[.09cm]
				Pure dephasing time at idle frequency $T_2^{\ast}$ & $0.826$ & $0.759$ & $0.25$ & \us \\[.09cm]
				Qubit-Qubit coupling $g_{j,j+1}/2\pi$ & $7.11$ & $7.20$ & $0.39$ & MHz \\[.09cm]
				Qubit-resonator coupling $g_{{\textrm{qr}}}/2\pi$ & $36.62$ & $38.15$ & $35.26$ & MHz \\[.09cm]
				Mean fidelity of single-qubit gates & $99.2$ & $99.4$ & $1.1$ & \% \\[.09cm]
			\end{tabular}
		\end{ruledtabular}
		\caption{Device parameters.}\label{tab:paras}
	\end{table*}
	
	\section{\label{sec:model} Model and Hamiltonian}
	\subsection{Pumping, polarization, and topology}
	
	One simple manifestation of topology in quantum systems is Thouless pumping, entailing quantized transport through an adiabatic cyclic evolution of a $1$D quantum system in the absence of the net electromagnetic field \cite{thouless_quantization_1983, xiao_berry_2010, citro_thouless_2023}. The physics and topological nature of pumping can be demonstrated by periodically modulating the Rice-Mele (RM) model  \cite{rice_elementary_1982} written as:
	\begin{equation}\label{eqRM}
		\hat{H}_{\textrm{RM}}=\sum_{j=0}^{L-2}[J+(-1)^j\delta](\hat{a}_j^{\dagger}\hat{a}_{j+1}+\mathrm{H.c.})+\sum_{j=0}^{L-1}(-1)^j\Delta\hat{a}_j^{\dagger}\hat{a}_j,
	\end{equation}
	where $\hat{a}^{\dagger}$($\hat{a}$) denotes the creation (annihilation) operator, the chain length $L=2N$ is even, $\pm\Delta$ are staggered on-site potentials, and $J\pm\delta$ are alternating hopping strengths.
	When $\Delta/2\pi=0$~MHz, the RM model in Eq.~\eqref{eqRM} reduces to the celebrated Su-Shrieffer-Heeger (SSH) model \cite{su_solitons_1979}.  Under periodic boundary conditions (PBC), the energy spectrum of $\hat{H}_{\rm RM}$ consists of two bands, separated by a gap except a gapless point at $(\Delta/2\pi, \delta/2\pi)=(0$~MHz$, 0$~MHz$)$.
	
	
	\begin{figure}[t]
		\centering
		\includegraphics[width=0.9\linewidth]{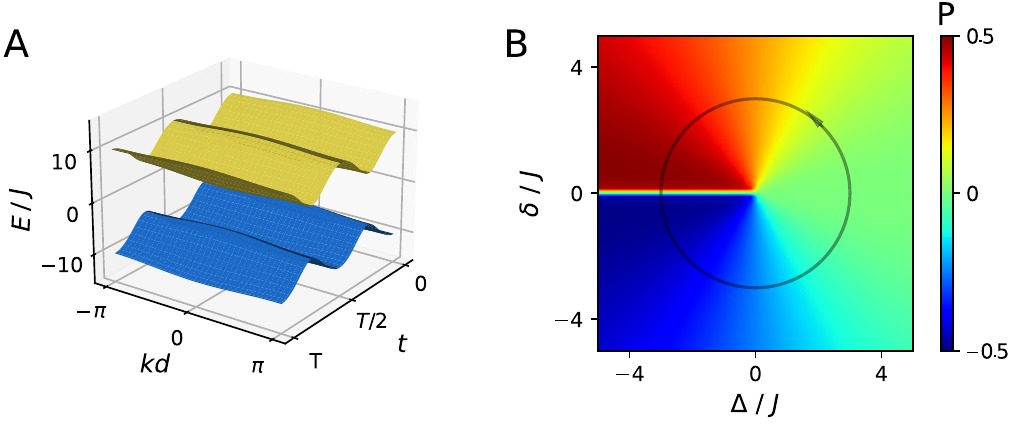}
		\caption{Energy bands of the Rice-Mele model and the polarization $\mathsf{P}$. (\textbf{A}) The band structure in the $k$-$t$ Brillouin zone under PBC, containing two bands separated by a gap. (\textbf{B}) Polarization $\mathsf{P}$ in a unit of the lattice constant ($d=2$) versus the Rice-Mele model parameters $\Delta$ and $\delta$.
			The discontinuity is due to the particular choice of the eigenstate phase.}
		\label{fig:pumpingbandpolar}
	\end{figure} 
	
	With PBC and periodic driving, we could define the Bloch wavefunction
	of the $n$-th band in a $k$-$t$ Brillouin zone as
	$\ket{\psi_{k,n}}=e^{\mathrm{i}kx}\ket{u_{n,k}}$, due to the periodicity
	of the lattice in both space and time.
	The band structure is plotted in Fig.~\ref{fig:pumpingbandpolar}A.
	When $\Delta$ and $\delta$ vary adiabatically with a sufficiently slow period  $T$ and the $n$-th band is evenly filled, we can define the Chern number as
	\begin{equation}\label{chern}
		\nu_n=\frac{1}{2\pi}\int_0^T\!\!\mathrm{d}t\int_{\mathrm{1BZ}}\!\!\!\!\mathrm{d}k\;\Omega_n(k, t),
	\end{equation}
	where
	$\displaystyle\Omega_n(k, t)\equiv\mathrm{i}(\bra{\partial_tu_{n,k}}\partial_ku_{n,k}\rangle-\bra{\partial_ku_{n,k}}\partial_tu_{n,t}\rangle)$ is the Berry curvature, and $\mathrm{1BZ}$ denotes the first Brillouin zone. The integer Chern number  characterizes quantized charge transport in the adiabatic limit, and the pumped amount of charge over one cycle $\Delta Q$ can be expressed as
	\begin{equation}
		\Delta Q=d\int_0^T\!\!\mathrm{d}t\;\bra{\psi(t)}\hat{\mathcal{J}}(t)\ket{\psi(t)},
	\end{equation}
	where $d=2$ is the lattice constant because of the staggered parameters. Here,
	\begin{equation}
		\hat{\mathcal{J}}(t)=\frac{\mathrm{i}}{L}\sum_{j=1}^L(J+(-1)^j\delta)\hat{a}_{j+1}^{\dagger}\hat{a}_{j}+\mathrm{H.c.}
	\end{equation} denotes the average current density \cite{wauters_localization_2019, wu_quantized_2022}, and $\ket{\psi(t)}$ is obtained by evolving the $N$-particle ground state $\ket{\psi(0)}$.
	
	
	By performing a Fourier transformation (FT) on the Bloch state, we can introduce the Wannier state localized at site $j$ for the $n$-th Bloch band as
	\begin{equation}\label{eqWannier}
		\ket{w_{n,j}}=\frac{1}{\sqrt{N}}\sum_k\mathrm{e}^{-\mathrm{i}jk}\ket{\psi_{n,k}}.
	\end{equation}

	The position matrix elements of the Wannier states in the thermodynamic limit ($N\to\infty$) can be calculated as
	\begin{equation}
		\begin{split}
			\mathsf{P}-j=\bra{w_{n,j}}\hat{x}-j\ket{w_{n,j}}=&\frac{1}{N}\sum_{k,k'}\bra{u_{n,k'}}\mathrm{e}^{-\mathrm{i}k'(x-j)}(x-j)\mathrm{e}^{\mathrm{i}k(x-j)}\ket{u_{n,k}}\\
			=& \frac{d^2}{4\pi^2}\int\!\mathrm{d}x\int\!\mathrm{d}k\mathrm{d}k'\;u^*_{n,k'}(x, t)\mathrm{e}^{-\mathrm{i}k'(x-j)}[-\mathrm{i}\partial_k\mathrm{e}^{\mathrm{i}k(x-j)}]u_{n,k}(x, t)\\
			=&\frac{d}{2\pi}\int_{\mathrm{1BZ}}\!\!\!\!{\mathrm{d}k}\;\bra{u_{n,k}}\mathrm{i}\partial_ku_{n,k}\rangle\\
			=&\frac{d}{2\pi}\int_{\mathrm{1BZ}}\!\!\!\!{\mathrm{d}k}\;\mathcal{A}_n^k(k, t),
		\end{split}
	\end{equation}
	where $\mathcal{A}_n^k(k, t)$ is the Berry connection, and $\mathsf{P}=\bra{w_{n,j}}\hat{x}\ket{w_{n,j}}$ denotes the polarization \cite{king-smith_theory_1993} as a quantum extension of the classical electric polarization. When the initial state is prepared as a Wannier state $\ket{w_{n,j}}$, the change of the polarization over one cycle is given by
	\begin{equation}
		\begin{split}
			\Delta \mathsf{P}=&\frac{d}{2\pi}\int_{0}^T\!\!\!\mathrm{d}t\int_{\mathrm{1BZ}}\!\!\!\!{\mathrm{d}k}\;\partial_t\mathcal{A}_n^k(k, t)\\
			=&\frac{d}{2\pi}\int_0^T\!\!\mathrm{d}t\int_{\mathrm{1BZ}}\!\!\!\!{\mathrm{d}k}\;(\mathrm{i}\partial_t\bra{u_{n,k}}\partial_ku_{n,k}\rangle-\mathrm{i}\partial_k\bra{u_{n,k}}\partial_tu_{n,k}\rangle)\\
			=&\nu_nd,
		\end{split}
	\end{equation}
	which can also be obtained by calculating the change of $\mathsf{P}$ for a parameter pair $(\Delta, \delta)$, as shown in Fig.~\ref{fig:pumpingbandpolar}B. The change of $\mathsf{P}$ is connected to the number of the evolving trajectory in $\Delta$-$\delta$ space,
	winding the gapless point $(0, 0)$.
	
	\subsection{Maximally localized Wannier state}
	\begin{figure}[t]
		\centering
		\includegraphics[width=0.97\linewidth]{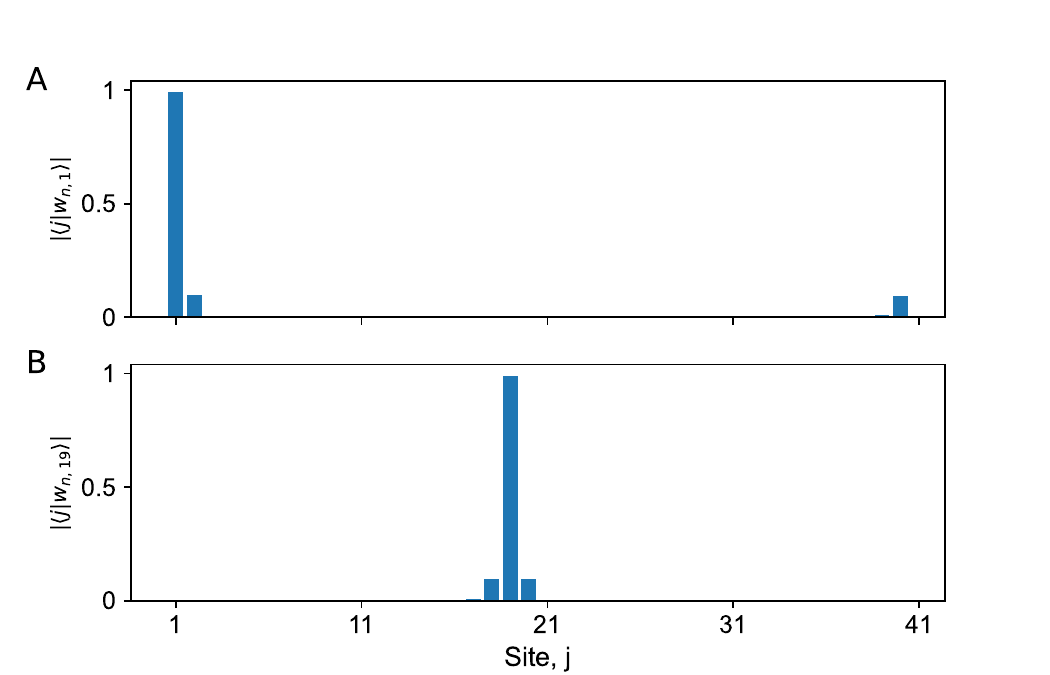}
		\caption{The wavefunctions of maximally localized Wannier state localized at the edge (\textbf{A}) and the center (\textbf{B}) of the Rice-Mele model with $N=20$.}
		\label{fig:wannierstate}
	\end{figure}
	Due to the gauge freedom of Bloch functions, the Wannier state defined in Eq.~\eqref{eqWannier} is non-unique \cite{marzari_maximally_2012}. More specifically, if we replace $\ket{\psi_{n,k}}$ to $\ket{\tilde{\psi}_{n,k}}=\mathrm{e}^{\mathrm{i}\varphi_n(k)}\ket{\psi_{n,k}}$, with $\varphi_n(k)$ being smooth real functions that is periodic in the momentum space, different sets of Wannier states will be defined, having distinct shapes and spreads. By introducing the second order moment of the position operator of $\ket{w_{n, 0}}$
	\begin{equation}
		\Omega=\sum_n\left[\bra{w_{n,0}}\hat{x}^2\ket{w_{n,0}}-\bra{w_{n,0}}\hat{x}\ket{w_{n,0}}^2\right]
	\end{equation}
	as the localization criterion, Marzari and Vanderbilt developed an effective
	approach to constrain the freedom gauge \cite{marzari_maximally_1997}. Through minimizing $\Omega$, we can obtain the maximally localized Wannier state (MLWS).
	
	For $1$D quantum systems, MLWS can be calculated as the eigenstates of the projected position operator in the real space, $\hat{P}\hat{x}\hat{P}$, with $\hat{P}=\sum\limits_{k}\ket{\psi_{n,k}}\bra{\psi_{n, k}}$ the projection operator to a filled band. However, the usual definition of $\hat{x}$ is vague when crossing the boundary under PBC. Alternatively, $\hat{x}$ can be substituted to a unitary operator $\hat{X}=\mathrm{e}^{\mathrm{i}\frac{2\pi}{N}\hat{x}}$, and the task to solve MLWS is transformed into diagonalization of 
	\begin{equation}
		\hat{X}_{P}=\hat{P}\hat{X}\hat{P}.
	\end{equation}
	Note that in general $\hat{X}_P$ is not Hermitian, meaning that the eigenstates are not orthogonal except for the thermodynamic limit.
	
		\begin{figure}[t]
		\centering
		\includegraphics[width=0.9\linewidth]{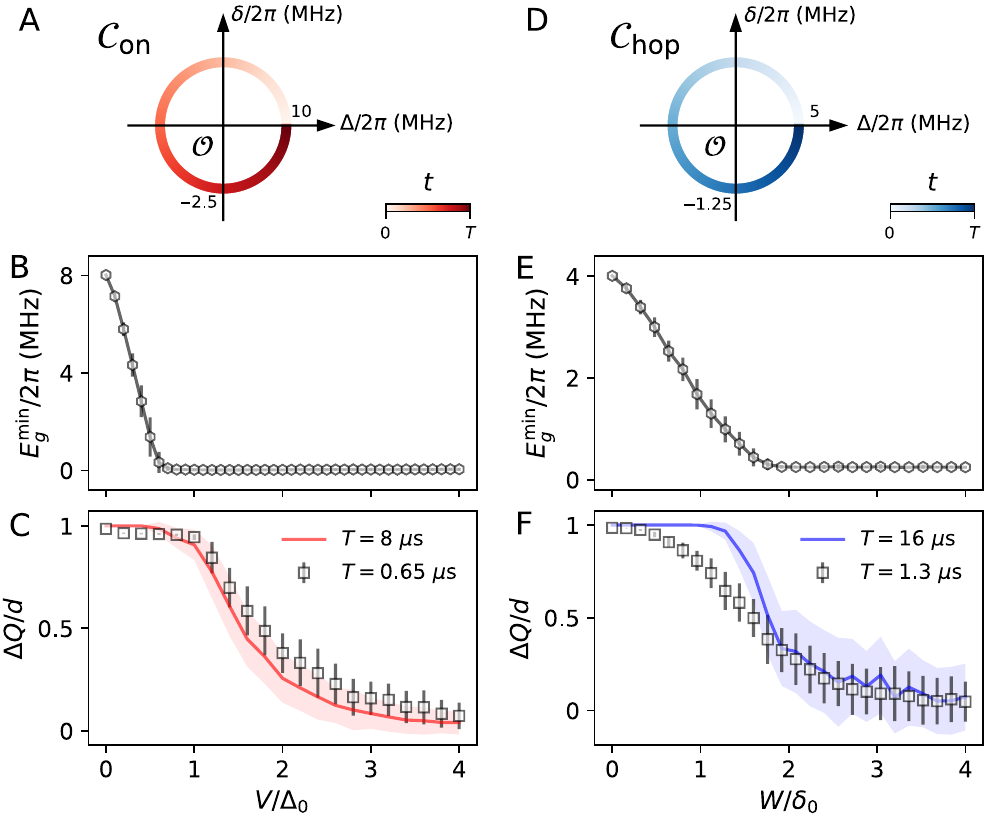}
		\caption{Bandgaps and transport in the Rice-Mele model with disorder.
			(\textbf{A}) Pumping trajectory $\mathcal{C}_{\mathrm{on}}$.
			(\textbf{B}) Minimum instantaneous gap, $E_g^{\mathrm{min}}$,
			versus the on-site random disorder strength $V$ for the trajectory $\mathcal{C}_{\mathrm{on}}$.
			(\textbf{C}) Pumped amount, $\Delta Q$, as a function of $V$ for the trajectory $\mathcal{C}_{\mathrm{on}}$.
			(\textbf{D}) Pumping trajectory $\mathcal{C}_{\mathrm{hop}}$.
			(\textbf{E}) Minimum instantaneous gap, $E_g^{\mathrm{min}}$, versus the hopping random disorder strength $W$ for the trajectory $\mathcal{C}_{\mathrm{hop}}$.
			(\textbf{F})  Pumped amount, $\Delta Q$, as a function of $W$ for the trajectory $\mathcal{C}_{\mathrm{hop}}$. All numerical simulations are averaged over 100 different disorder configurations, and the shaded regions and the error bars show the one standard deviation (1SD).}
		\label{fig:simulationdisordergapchern}
	\end{figure}
	
	Numerically, we calculate MLWS of the RM model for each energy band based on the experimental setup. Two typical wavefunctions of MLWS localized at the edge and center individually are shown in Fig.~\ref{fig:wannierstate}. 
	We find that a single particle localized at site $j$ approximates a Wannier state of a
	half-filling (upper or lower) band that is determined by the parity of the site \cite{cerjan_thouless_2020}. The quantum state fidelity between a single-particle excitation and the corresponding MLWS is larger than $0.99$.
	
	Thus, quantized transport, indicated by the Chern number, can be observed experimentally by measuring the displacement of the center-of-mass (CoM) per period, $\displaystyle\delta x =\bar{x}(T)-\bar{x}(0)$, during the adiabatic evolution after preparing a single-excitation initial state \cite{nakajima_topological_2016, lohse_thouless_2016, ke_multiparticle_2017}. The CoM, expressed as
	\begin{align}
		\bar{x}=\sum_jj\langle \hat{n}_j\rangle,
	\end{align}
	with $\hat{n}_j\equiv\hat{a}_j^\dag\hat{a}_j$,
	can be extracted from the adiabatic evolution of the initial single-excitation state on the superconducting processor.

	
	
	\subsection{Pumping under disorder}\label{sec:disorder}
	In addition to the topological nature of Thouless pumping, we investigate the effects of on-site random disorder $V_j$ on the topological pumping by substituting $\Delta$ in Eq.~\eqref{eqRM} to $\Delta+V_j$,
	where $V_j$ is uniformly distributed in the range $[-V, V]$.
	For an integrable 1D quantum system, an arbitrarily small amount of on-site disorder
	leads to localization \cite{anderson_absence_1958}.
	However, robustness against weak disorder in topological quantum phenomena is expected \cite{anirban_15_2023}. Here, we numerically study the topology-disorder transition, in our superconducting processor.
	
	We now consider the trajectory $\mathcal{C}_{\mathrm{on}}$  (Fig.~\ref{fig:simulationdisordergapchern}A),
	\begin{align}
		(\Delta, \delta)=(\Delta_0\cos(2\pi t/T), \delta_0\sin(2\pi t/T)),
	\end{align}
	with $\Delta_0/2\pi=$\SI{10.0}{\MHz}, $\delta_0/2\pi=$\SI{2.5}{\MHz}, $J/2\pi=$\SI{2.0}{\MHz}, 
	and $T=0.65$~$\mu$s. We calculate the average of $\Delta Q$ for different values of $V$, for
	$T=0.65$~$\mu$s and $8.0$~$\mu$s. As shown in Fig.~\ref{fig:simulationdisordergapchern}C, they both satisfy the adiabatic
	limit.  Moreover, the
	crossover between the topological phase and the localized phase is related to the
	minimal many-body instantaneous gap \cite{citro_thouless_2023, wauters_localization_2019} (Fig.~\ref{fig:simulationdisordergapchern}B)
	\begin{align}
		E_g^{\mathrm{min}}=\min_{\phi}[E_{N+1}(t)-E_N(t)],
	\end{align}
	where $E_{N}(t)$ is the $N$-particle ground state energy at time $t$.
	We observe that topological pumping persists with $\Delta Q/d=1$, even with weak disorder
	$V/\Delta_0\approx 1$, when the energy gap remains open to allow for
	a possible adiabatic evolution of the ground state, and
	$\Delta Q/d\approx 0$ with strong disorder satisfying $V/\Delta_0\gtrsim 3$.
	
	Moreover, we study the effect of hopping random disorder by substituting $\delta$ by $\delta + W_j$ in Eq.~(\ref{eqRM}) by taking $W_j$ being uniformly distributed in the range $[-W, W]$. The trajectory $\mathcal{C}_{\mathrm{hop}}$ is chosen as an equally scaled-down
	version of $\mathcal{C}_{\mathrm{on}}$, with $\Delta_0/2\pi={5.0}$~{MHz}, $\delta_0/2\pi={1.25}$~{MHz}, $J/2\pi={1.0}$~{MHz}, and $T=1.3$~$\mu$s, see Fig.~\ref{fig:simulationdisordergapchern}D.
	We observe in Fig.~\ref{fig:simulationdisordergapchern}F a decreasing behavior of $\Delta Q$ as the hopping disorder strength increases, which is similar to the case with on-site random disorder. Note that the non-adiabatic effect shows that the transition point moves toward a weaker disorder strength, by comparing the cases with periods $T=1.3$~$\mu$s and $16$~$\mu$s. 

	\subsection{Localization in the Rice-Mele model with disorder}
	\begin{figure}
		\centering
		\includegraphics[width=0.9\linewidth]{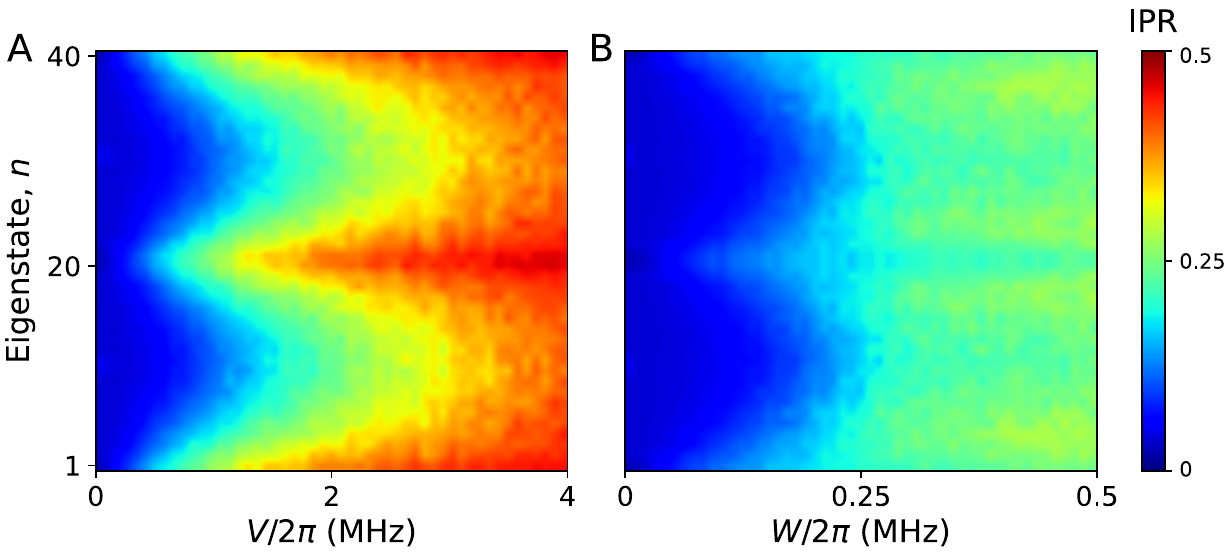}
		\caption{The inverse participation ratio (IPR) numerically calculated for the RM model with on-site and hopping disorder. (\textbf{A}) IPR of the $n$-th eigenstate against the on-site random disorder strength $V$ for the trajectory $\mathcal{C}_{\mathrm{on}}$ at the SSH parameter point $(\Delta/2\pi, \delta/2\pi)=(\SI{0}{\MHz}, \SI{2.5}{\MHz})$.
			(\textbf{B}) IPR of the $n$-th eigenstate against the hopping random strength $W$ for the trajectory $\mathcal{C}_{\mathrm{hop}}$ at the parameter point $(\Delta/2\pi, \delta/2\pi)=(\SI{5}{\MHz}, \SI{0}{\MHz})$. Both results are averaged with $100$ different configurations of disorder.}
		\label{fig:onhopipr}
	\end{figure}
	
	According to Anderson localization (AL), an integrable $1$D system tends to be at a localized state with relatively strong on-site disorder.
	To characterize the localization phenomena of the RM model with disorder,
	we employ the real-space inverse participation ratio (IPR), which is
	defined as
	\begin{equation}
		\mathcal{I}(\ket{\psi_n})=\sum_j|\bra{1_j}\psi_n\rangle|^4,
	\end{equation}
	where $\ket{\psi_n}$ is the $n$-th eigenstate, and $\ket{1_j}=\hat{a}_j^{\dagger}\ket{0}$ denotes a particle state with an excitation at site $j$. A large IPR signifies a strong localization tendency, and $\mathcal{I}\sim 1/L$ for a plane-wave state,
	while $\mathcal{I}=1$ identifies a perfect localized state at the single site.
	
	We calculate the IPR of the system with on-site random disorder
	at the so-called SSH parameter point, i.e.,
	$\displaystyle (\Delta/2\pi, \delta/2\pi)=(\SI{0}{\MHz}, \SI{2.5}{\MHz})$ of the trajectory $\mathcal{C}_{\mathrm{on}}$.
	Figure~\ref{fig:onhopipr}A clearly shows that the system tends to be at a localized state as the disorder strength becomes large. Note that the transition point between the topological phase and the localized phase, $V_{\mathrm{c}}\simeq 0.2$~MHz, is much smaller than the trajectory radius $\Delta_0/2\pi=10$~MHz. Although the instantaneous Hamiltonian eigenstates are shown in Fig.~\ref{fig:onhopipr}A to be localized with a small on-site disorder strength, the topological transport remains robust against weak disorder. It was shown \cite{wauters_localization_2019} that the breakdown of pumping under on-site disorder is linked to the delocalization-localization transition of the single-particle 
	Floquet eigenstates instead of instantaneous eigenstates.
	

	Similar to the case with on-site random disorder, an integrable $1$D system tends to be localized under considerable hopping random disorder, but with a distinct law \cite{theodorou_extended_1976, fleishman_fluctuations_1977, soukoulis_off-diagonal_1981}. When only hopping random disorder is involved, the localization length of the zero-energy state is infinite. Nevertheless, the state should be considered to be localized due to the fact that the mean values of the transmission coefficient approach zero in the thermodynamic limit.

	In analogy to the SSH parameter point, we compute the IPR for the trajectory $\mathcal{C}_{\mathrm{hop}}$ with hopping disorder at the parameter point  $(\Delta/2\pi, \delta/2\pi)=(\SI{5}{\MHz}, \SI{0}{\MHz})$, see Fig.~\ref{fig:onhopipr}B. The numerical results show that the IPR increases as the disorder strength increases, suggesting that instantaneous eigenstates tend to be localized with relatively strong disorder.
	Similarly, the transition point of instantaneous eigenstates is much smaller than the breakdown point of pumping. Therefore, it is still unclear whether the behavior of $\Delta Q$ could characterize the delocalization-localization transition of the single-particle Floquet eigenstates, which deserves further investigations.
	
		\begin{figure}[t]
		\centering
		\includegraphics[width=0.9\linewidth]{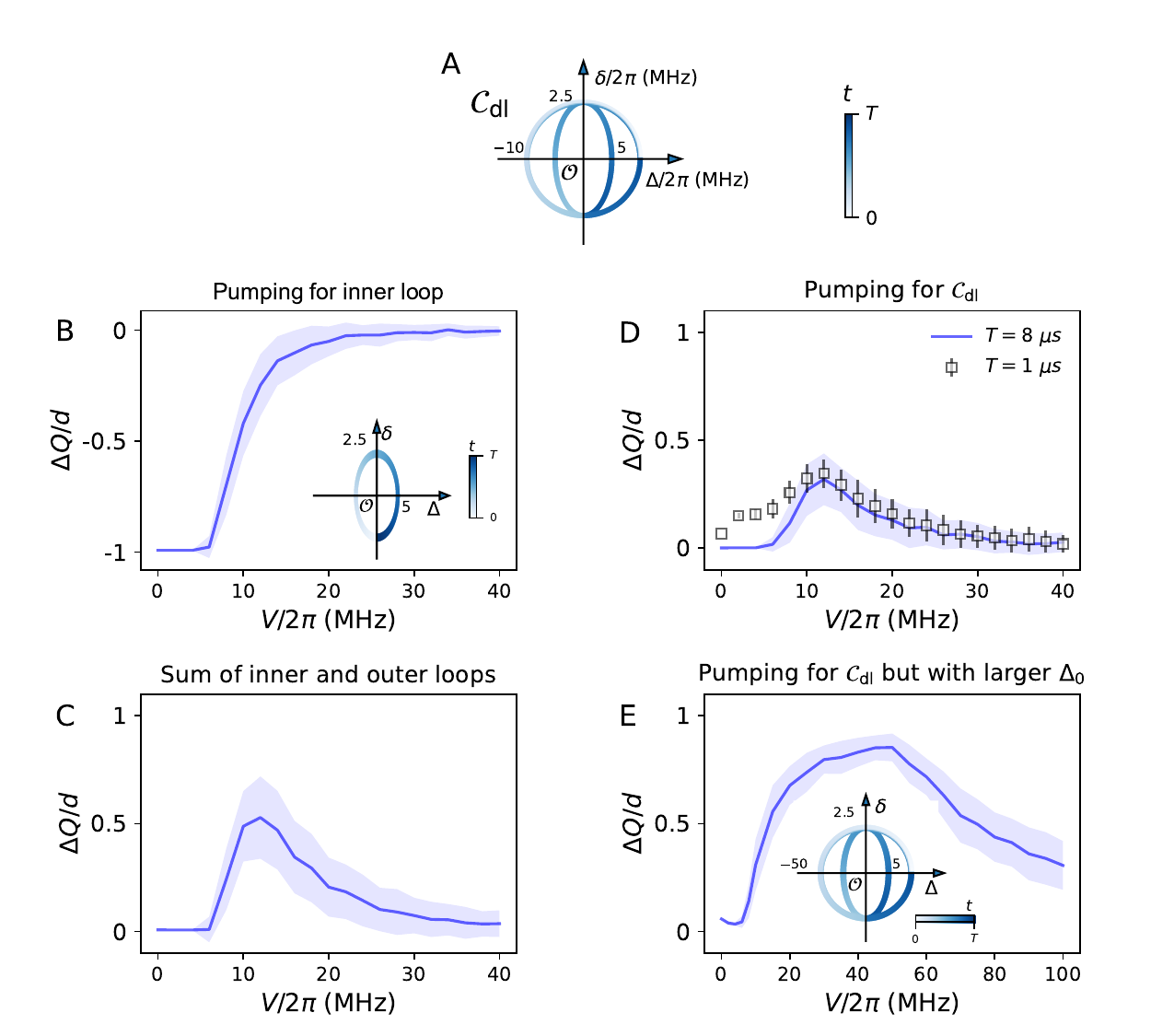}
		\caption{Topological pumping with on-site random disorder for the double-loop trajectory, $\mathcal{C}_{\mathrm{dl}}$.
			(\textbf{A}) Trajectory $\mathcal{C}_{\mathrm{dl}}$, containing an incomplete counterclockwise outer loop, and a complete clockwise inner loop. (\textbf{B}) Pumped amount, $\Delta Q$, versus the disorder strength $V$ for the inner loop. (\textbf{C}) $\Delta Q$ versus $V$ for the complete outer loop, corresponding to the result in
			Fig.~\ref{fig:simulationdisordergapchern}\textbf{C}, and $\Delta Q$ versus $V$ for the inner loop.  (\textbf{D}) $\Delta Q$ versus $V$ for   $\mathcal{C}_{\mathrm{dl}}$ with $T=1$~$\mu$s and $8$~$\mu$s, respectively.
			(\textbf{E}) Quantized topological pumping induced by on-site random disorder with sufficiently separated $\Delta_0$ and $\Delta_1$, where $\Delta_0/2\pi=50$~MHz, and $\Delta_1/2\pi=5$~MHz.}
		\label{fig:simulationodicpchern}
	\end{figure}
	
	\subsection{Double-loop pumping induced by on-site disorder}
	Connecting two elliptical loops with opposite directions,
	the double-loop trajectory, $\mathcal{C}_{\mathrm{dl}}$,
	as shown in Fig.~\ref{fig:simulationodicpchern}{A},
	is topologically trivial in the clean limit;
	because of the addition of two inverse Chern numbers
	\cite{wu_quantized_2022, nakajima_competition_2021}.
	Topological pumping would be realizable, if disorder asymmetrically inhibits the topology of the two pumping loops.

	Here, we now study a double-loop trajectory consisting of one two half outer loops $\mathcal{C}_{\mathrm{on}}$ and one complete inter loop $\mathcal{C}_{\mathrm{on}1}$ in an inverse direction, as shown in Fig.~\ref{fig:simulationodicpchern}\textbf{A}, which is expected to
	exhibit similar topological transport behaviors with on-site random disorder, similar as the case discussed in Ref.~\cite{nakajima_competition_2021}.

	The trajectory, $\mathcal{C}_{\mathrm{dl}}$, can be parameterized as $(\Delta(t), \delta(t))$  with
	\begin{equation}\label{eqDelta}
		\Delta(t)=
		\begin{cases}
			\Delta_0\cos\Omega_1t, & \textrm{for~~}0<t\leq 3\tau_1/4,\\
			-\Delta_1\sin\Omega_2(t-3\tau_1/4), & \textrm{for~~}3\tau_1/4<t\leq 3\tau_1/4+\tau_2,\\
			\Delta_0\sin\Omega_1 (t-3\tau_1/4-\tau_2),&\textrm{for~~}3\tau_1/4+\tau_2<t\leq\tau_1+\tau_2,
		\end{cases}
	\end{equation}
	\begin{equation}\label{eqdelta}
		\delta(t)=
		\begin{cases}
			\delta_0\sin\Omega_1t, & \textrm{for~~}0<t\leq 3\tau_1/4,\\
			-\delta_0\cos\Omega_2(t-3\tau_1/4), & \textrm{for~~}3\tau_1/4<t\leq 3\tau_1/4+\tau_2,\\
			-\delta_0\cos\Omega_1(t-3\tau_1/4-\tau_2), & \textrm{for~~}3\tau_1/4+\tau_2<t\leq\tau_1+\tau_2,
		\end{cases}
	\end{equation}
	with $\Delta_0/2\pi=10$~MHz, $\Delta_1/2\pi=5$~MHz, $\delta_0/2\pi=2.5$~MHz, $J/2\pi=2$~MHz, $\Omega_1=2\pi/\tau_1$, $\Omega_2=2\pi/\tau_2$, and $T=2\tau_1=2\tau_2=1$~$\mu$s.
	We show $\Delta Q$ of $\mathcal{C}_{\mathrm{on}1}$ versus the on-site disorder strength in Fig.~\ref{fig:simulationodicpchern}B and a summation of $\Delta Q$ for $\mathcal{C}_{\mathrm{on}}$ and $\mathcal{C}_{\mathrm{on}1}$ in Fig.~\ref{fig:simulationodicpchern}C. These numerical results demonstrate topological pumping induced by moderate on-site disorder. In addition, we directly calculate $\Delta Q$ of $\mathcal{C}_{\mathrm{dl}}$ versus the disorder strength, which matches well with the summation of $\Delta Q$ for two loops, see Fig.~\ref{fig:simulationodicpchern}D.
	
	Quantized topological pumping occurs when $\Delta_0$ and $\Delta_1$ are sufficiently far apart, such that an appropriate disorder strength $V$ is strong enough for the inner loop $\mathcal{C}_{\mathrm{on}1}$, but still too weak for the outer loop $\mathcal{C}_{\mathrm{on}}$. We numerically simulate another trajectory with identical parameters to $\mathcal{C}_{\mathrm{dl}}$ except for
	$\Delta_0/2\pi=50$~MHz, where a quantized plateau is observed, see Fig.~\ref{fig:simulationodicpchern}E.

	\subsection{Single-loop pumping induced by hopping disorder}\label{section:sl}
	
\begin{figure}[t]
		\centering
		\includegraphics[width=0.9\linewidth]{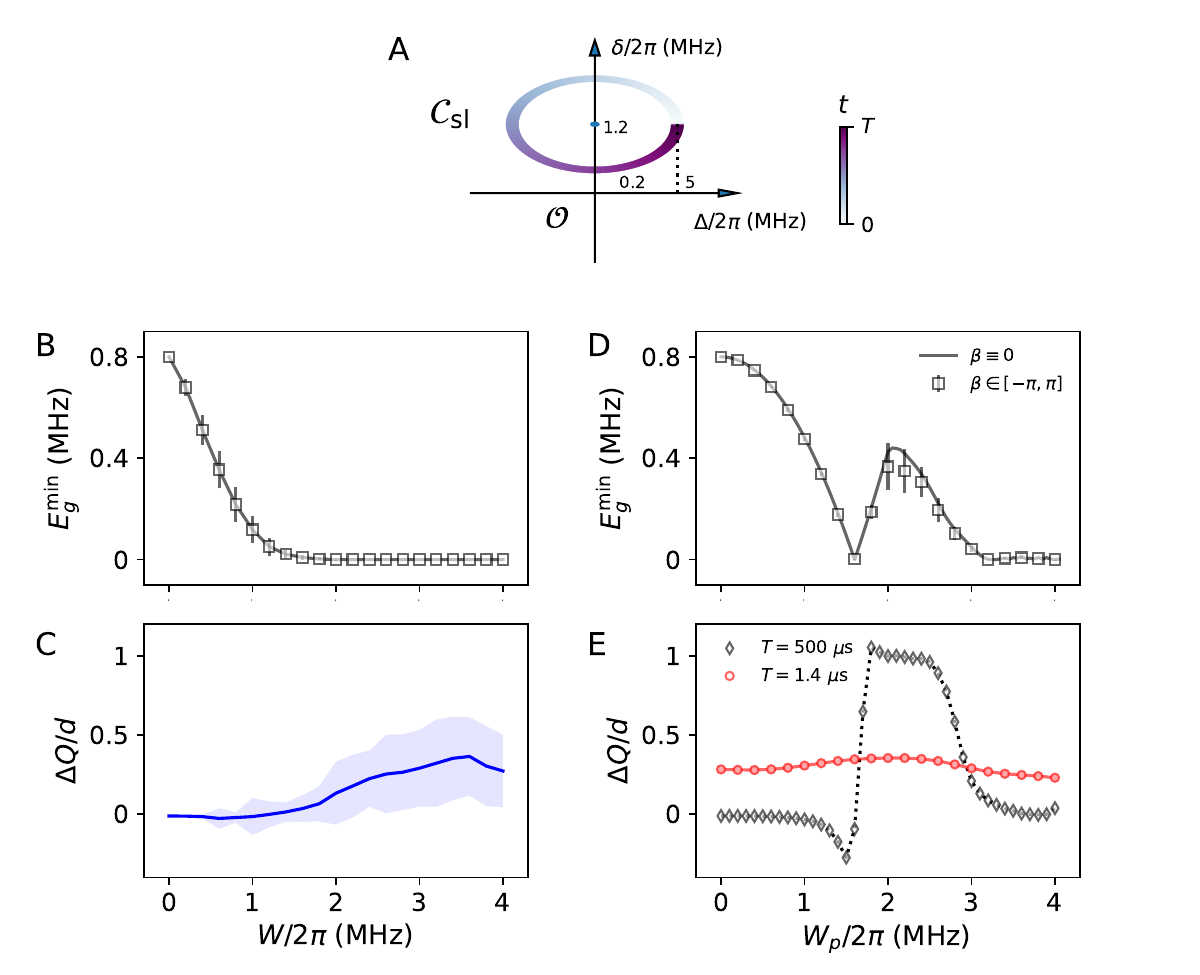}
		\caption{
			Topological pumping with hopping random disorder and  quasi-periodic intracell hopping disorder for the single-loop trajectory, $\mathcal{C}_{\mathrm{sl}}$. (\textbf{A}) The single-loop trajectory $\mathcal{C}_{\mathrm{sl}}$, containing a circle centered at the point $(0,\delta_c)$ on the $\delta$-axis and being outside  of the origin in the $\Delta$-$\delta$ plane.
			(\textbf{B}) The minimum instantaneous gap, $E_g^{\mathrm{min}}$, versus hopping random disorder strengths $W$.
			(\textbf{C}) The pumped amount, $\Delta Q$, against $W$ for $T=500$~$\mu$s. The calculation result of $\Delta Q$ is averaged over 100 different disorder samples, and the shaded regions and the error bars show the range of one standard deviation.
			(\textbf{D}) $E_g^{\mathrm{min}}$ versus the quasi-periodic  intracell hopping disorder strength $W_p$ with the phase $\beta$ fixed at $0$ and taken from the uniform distribution in the range $[-\pi, \pi]$, respectively.
			(\textbf{E}) $\Delta Q$ against $W_p$ for $T=1.4$~$\mu$s and $500$~$\mu$s, showing a quantized plateau. The lattice size is $L=800$ for the numerical calculation of the band gap. 
		}
		\label{fig:simulationhtatpchern}
	\end{figure}

	In the clean limit, the trajectory with a circle outside the origin centered on the $\delta$-axis, $\mathcal{C}_{\mathrm{sl}}$,
	as shown in Fig.~\ref{fig:simulationhtatpchern}A, leads to a topologically trivial pumping phenomenon with a winding number of $0$. By introducing an appropriate hopping random disorder strength, a nonzero $\Delta Q$  occurs, because the gapless point (line) possibly moves inside the trajectory. However, pumping is not quantized, precluding the adiabatic limit, due to the rapid closure of the  band gap. Fortunately, an intrinsic topologically nontrivial quantized pumping scheme, induced by quasi-periodic intracell hopping  disorder with gap reopening, has been proposed in Ref.~\cite{wu_quantized_2022}.
	
\begin{figure}[t]
		\centering
		\includegraphics[width=0.7\linewidth]{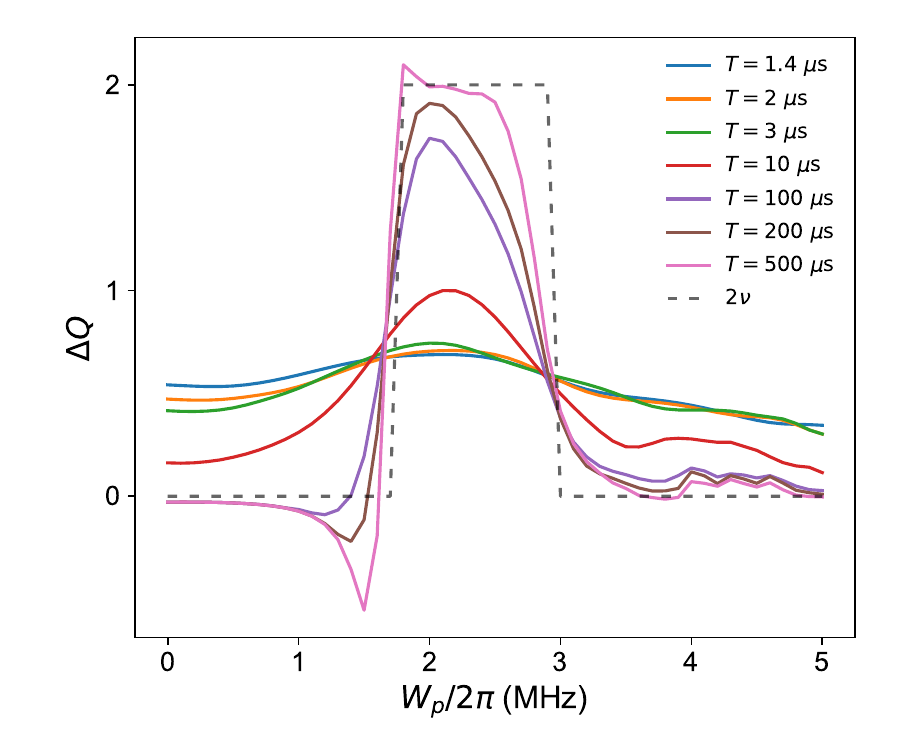}
		\caption{Pumping for the trajectory, $\mathcal{C}_{\mathrm{sl}}$, with quasi-periodic intracell hopping disorder for different periods. The dashed line in the Chern number obtained by numerical results. The quantized plateau appears when the disorder strength, $W_p/2\pi$, varies between around $1.6$~MHz and $2.9$~MHz, indicated by two dashed-dotted lines.}
		\label{fig:hophtatpperiod}
	\end{figure}

\begin{figure}[]
		\centering
		\includegraphics[width=0.97\linewidth]{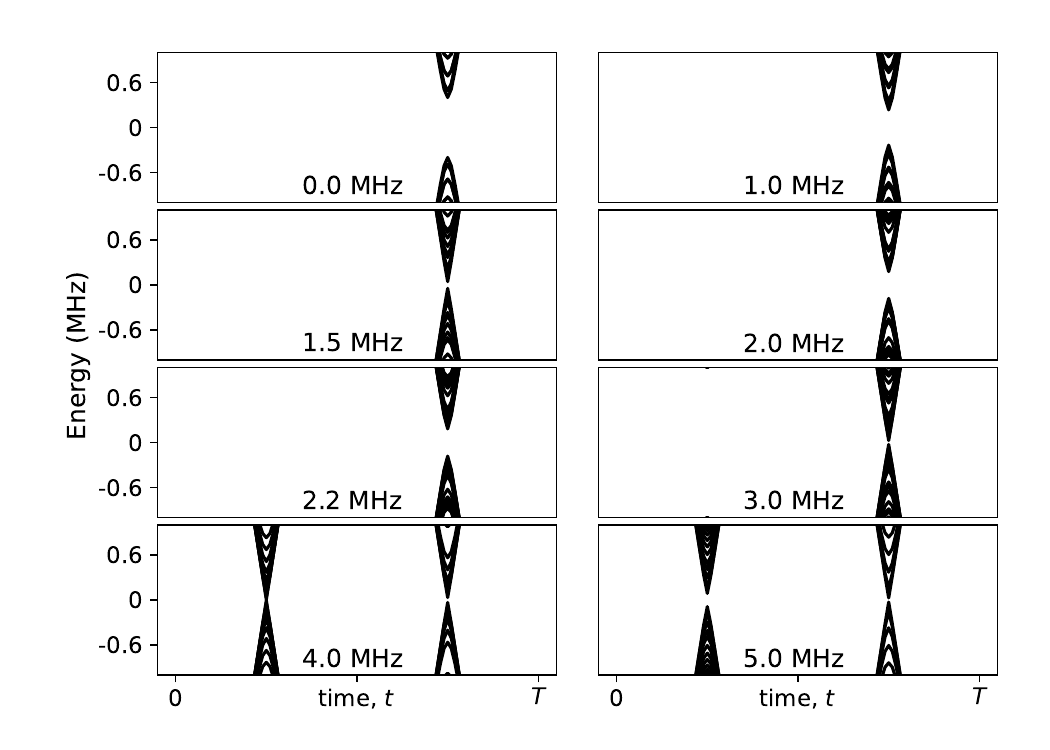}
		\caption{Instantaneous energy spectra of the bulk under quasi-periodic disorder. The results are averaged over 100 disorder realizations.}
		\label{fig:htatpspectrum}
	\end{figure}

	Given the parameterized path, $\mathcal{C}_{\mathrm{sl}}$, parameterized with $(\Delta, \delta)=(\Delta_0\cos\phi, \delta_c+\delta_0\sin\phi)$, with $\Delta_0/2\pi=5$~MHz, $\delta_0/2\pi=1.0$~MHz, $\delta_c/2\pi=1.2$~MHz, and $J/2\pi=1.8$~MHz,
	we calculate $E_g^{\mathrm{min}}$ and $\Delta Q$ against random hopping disorder $W_j\in[-W, W]$, and quasi-periodic intracell hopping disorder
	\begin{align}
		W_{j=2k}'=W_p\cos(2\pi\alpha k+\beta),
	\end{align}
	where $\alpha=(\sqrt{5}-1)/2$ is an irrational number.
	Although $\Delta Q$ becomes nonzero with an intermediate random disorder strength $W/2\pi\gtrsim2$~MHz, the minimum gap closes, resulting in a non-quantized pumping with large sample-to-sample fluctuations, see Figs.~\ref{fig:simulationhtatpchern}B and \ref{fig:simulationhtatpchern}C. In comparison, the gap can reopen with quasi-periodic hopping disorder, and one quantized plateau is observed at {1.5~MHz} $\lesssim W_p/2\pi\lesssim 3$~MHz, see Fig.~\ref{fig:simulationhtatpchern}D and \ref{fig:simulationhtatpchern}E. The spectra under different quasiperiodic disorder is shown in Fig.~\ref{fig:htatpspectrum}, which can be also observed by dynamical spectrum technique \cite{shi2022observing}. The increase of $\Delta Q$ with a short period $T=1.4$~$\mu$s provides an opportunity for an experimental demonstration of Thouless pumping induced by quasi-periodic disorder, as a dynamical analog of topological Anderson insulator (TAI) in Ref.~\cite{wu_quantized_2022}. To exactly obtain the range of disorder, where the plateau appears, we plot $\Delta Q$ for different  disorder strengths and periods, see Fig.~\ref{fig:hophtatpperiod}. The numerical results indicate that the quantized plateau can be observed, when {1.6~MHz} $\leq W_p/2\pi\leq 2.9$~{MHz}, in the adiabatic limit. Although the maximum of $\Delta Q$ decreases with the descent of period, the peak between the parameter region indicated by the Chern number, i.e., the dashed line in Fig.~\ref{fig:hophtatpperiod}, retains.
	
	\clearpage
	
	\clearpage
	\section{Floquet Engineering  for Adiabatic Systems}\label{sec:zfield}
	Floquet engineering has been applied in superconducting quantum circuits as an effective method to modulate the hopping strength of a time-independent quantum many-body system \cite{zhao_probing_2022, shi2022observing}.
	In our 1D 41-qubit superconducting processor, the effective Hamiltonian of two nearest-neighboring qubits can be described as
	\begin{equation}\label{eqTwoQubit}
		\hat{H}=-\omega_j(t)\hat{\sigma}_j^z/2-\omega_{j+1}(t)\hat{\sigma}_{j+1}^z/2+g_{j,j+1}(\hat{\sigma}_j^+\hat{\sigma}_{j+1}^-+\mathrm{H.c.}),
	\end{equation}
	where $\omega_j(t)$ denotes the $j$-th qubit frequency, $\hat{\sigma}_j^{x,y,z}$ are Pauli matrices with $\hat{\sigma}_j^{\pm}=(\hat{\sigma}_j^{x}\pm\mathrm{i}\hat{\sigma}_j^{y})/2$, $g_{j,j+1}$ is the coupling strength between the $j$-th and $(j+1)$-th qubits. With the method introduced in Ref.~\cite{shi2022observing}, the effective hopping strength can be modulated as
	\begin{equation}
		g^{\textrm{eff}}_{j,j+1}=g_{j,j+1}J_0\left(\frac{\eta_jA_j-\eta_{j+1}A_{j+1}}{\mu}\right),
	\end{equation}
	when we apply a microwave to manipulate the qubit frequency as
	\begin{equation}\label{eqMicrowave}
		\omega_j(t)=\bar{\omega} + A_j\sin(\mu t+\varphi_0),
	\end{equation}
	where $A_j$ and $\mu$ denote the modulation amplitude and the frequency, respectively, $\bar{\omega}$ is the average qubit frequency, $\varphi_0$ is a common initial phase, $J_s(x)$ is the $s$-order Bessel function, and $\eta_j\approx 1$ is the scale factor corresponding to the experimental calibration.
	
	\subsection{Adiabatic condition}
	In the Rice-Mele model Hamiltonian \eqref{eqRM}, the simultaneous adiabatic changes of the on-site and hopping terms result in a two-qubit Hamiltonian:
	\begin{equation}\label{eqHam1}
		\hat{H}'=\Delta_j(t)\hat{n}_{j}+\Delta_{j+1}(t)\hat{n}_{j+1}+\delta_{j,j+1}(t)(\hat{a}_{j}^{\dagger}\hat{a}_{j+1}+\hat{a}_{j}\hat{a}_{j+1}^{\dagger}).
	\end{equation}
	Then the Hamiltonian in the interaction picture can be expressed as
	\begin{equation}\label{eqInterPic1}
		\hat{H}_I'=\delta_{j,j+1}(t)\exp\left\{\mathrm{i}\int_0^t\!\mathrm{d}t\;\Delta(t)\right\}\hat{a}_j^{\dagger}\hat{a}_{j+1}+\mathrm{H.c.},
	\end{equation}
	where  $\Delta(t)\equiv\Delta_j(t)-\Delta_{j+1}(t)$. Intuitively, if $\delta_j(t)$ and $\Delta_j(t)$ vary slowly enough, we could generalize Eq.~\eqref{eqMicrowave} to
	\begin{equation}\label{eq:freq_wave}
		\omega_j(t)=\bar{\omega}+\Delta_j(t)+A_j(t)\sin(\mu t+\varphi_0),
	\end{equation}
	where
	\begin{equation}
		A_{j}(t)=\frac{1}{\eta_j}\left[\eta_{j+1}A_{j+1}(t)\mp\mu J_0^{-1}\left(\frac{\delta_{j,j+1}(t)}{g_{j,j+1}}\right)\right].
	\end{equation}
	More specifically, if we denote the cut-off frequency of $\delta_j(t)$ and $\Delta_j(t)$ as $\delta_j^c$ and $\Delta_j^c$, the adiabatic condition will be expressed as
	\begin{equation}\label{eq:adia_cond}
		\delta_j^c,\Delta_j^c\ll \mu.
	\end{equation}
	
\subsection{Nyquist condition}

	\begin{figure}[t]
	\centering
	\includegraphics[width=0.97\linewidth]{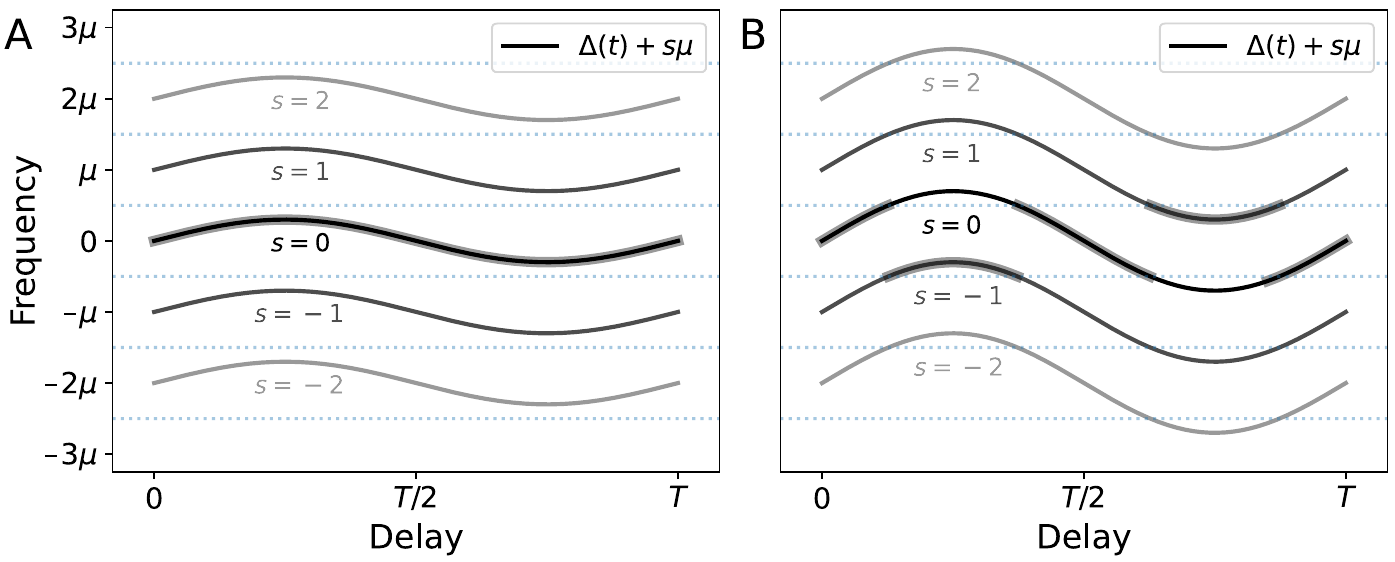}
	\caption{Bands of frequencies  inside \textbf{(A)} and outside  \textbf{(B)} the Nyquist condition. The thick solid lines represent the effective band under the rotating wave approximation (RWA). The lowest-order term remains $\Delta(t)$, when $\mu>2\max\limits_t|\Delta(t)|$ is satisfied; otherwise, the effective band consists of parts of the bands $s=\pm 1$.}
	\label{fig:nyquist}
\end{figure}

When the qubit frequency  is adjusted to be as in Eq.~\eqref{eq:freq_wave}, the unitary transformation in the interaction picture is written as
\begin{equation}
	\hat{U}_I(t)\approx\prod_{n=j,j+1}\exp\left\{\mathrm{i}\left(\bar{\omega}t+\int_0^t\!\mathrm{d}t\;\left[\Delta_n(t)-\frac{A_n(t)}{\mu}\cos(\mu t+\varphi_0)\right]\right)
		\hat{n}_j\right\},
\end{equation}

and  the effective Hamiltonian can be obtained as
\begin{equation}\label{eqInterPic2}
	\begin{split}
		\hat{H}_I'=&\hat{U}_I(t)\hat{H}(t)\hat{U}_I^{\dagger}(t)+\mathrm{i}\left[\frac{\mathrm{d}}{\mathrm{d}t}\hat{U}_I(t)\right]\hat{U}_I^{\dagger}(t),\\
		=&g_{j,j+1}\exp\left\{\mathrm{i}\int_0^t\!\!\mathrm{d}t\;\Delta(t)\right\}\exp\left\{\mathrm{i}\frac{A_{j}(t)}{\mu}\cos(\mu t+\varphi_{0})\right\}\exp\left\{-\mathrm{i}\frac{A_{j+1}(t)}{\mu}\cos(\mu t+\varphi_{0})\right\}\hat{a}_j^+\hat{a}_{j+1}+\mathrm{H.c.}\\
		=&g_{j,j+1}\exp\left\{\mathrm{i}\int_0^t\!\!\mathrm{d}t\;\Delta(t)\right\}\sum_{m,n=-\infty}^{+\infty}\!\mathrm{i}^{m+n}J_m\left(\frac{A_j(t)}{\mu}\right)J_n\left(-\frac{A_{j+1}(t)}{\mu}\right)\mathrm{e}^{\mathrm{i}(m+n)(\mu t+\varphi_0)}\hat{a}_j^+\hat{a}_{j+1}+\mathrm{H.c.}\\
		=&g_{j,j+1}\exp\left\{\mathrm{i}\int_0^t\!\!\mathrm{d}t\;\Delta(t)\right\}\sum_{s=-\infty}^{+\infty}\sum_{m+n=s}\mathrm{i}^{m+n}J_m\left(\frac{A_j(t)}{\mu}\right)J_n\left(-\frac{A_{j+1}(t)}{\mu}\right)\mathrm{e}^{\mathrm{i}(m+n)(\mu t+\varphi_0)}\hat{a}_j^+\hat{a}_{j+1}+\mathrm{H.c.}\\
		=&g_{j,j+1}\exp\left\{\mathrm{i}\int_0^t\!\!\mathrm{d}t\;\Delta(t)\right\}\sum_{s=-\infty}^{+\infty}\exp\left\{\mathrm{i}s(\mu t+\frac{\pi}{2}+\varphi_0)\right\}J_s\left(\frac{A_{j+1}(t)-A_j(t)}{\mu}\right)\hat{a}_j^+\hat{a}_{j+1}+\mathrm{H.c.}.
	\end{split}
\end{equation}
Compared with Eqs.~\eqref{eqInterPic1}, Eq.~\eqref{eqInterPic2} indicates that the Hamiltonian is composed of bands of frequencies, i.e., $\Delta(t) + s\mu$. According to the rotating wave approximation (RWA), high-frequency oscillatory terms are neglected, and low-frequency terms contribute predominantly. The necessary and sufficient condition for frequency band components to have no overlap with each other in the frequency domain is
\begin{equation}\label{eqNyquistCond}
	\mu > 2\max_t|\Delta(t)|,
\end{equation}
as shown in Figs.~\ref{fig:nyquist}A and \ref{fig:nyquist}B. Inequality \eqref{eqNyquistCond} is called the Nyquist condition, named after the Nyquist's sampling theorem \cite{marks_introduction_1991}. When both Eqs.~\eqref{eq:adia_cond} and \eqref{eqNyquistCond} are satisfied, Eq.~\eqref{eqInterPic2} approximates to Eq.~\eqref{eqInterPic1}.


\section{Additional Experimental Data}
\subsection{Effects of decoherence}

\begin{figure}[t]
	\centering
	\includegraphics[width=0.55\linewidth]{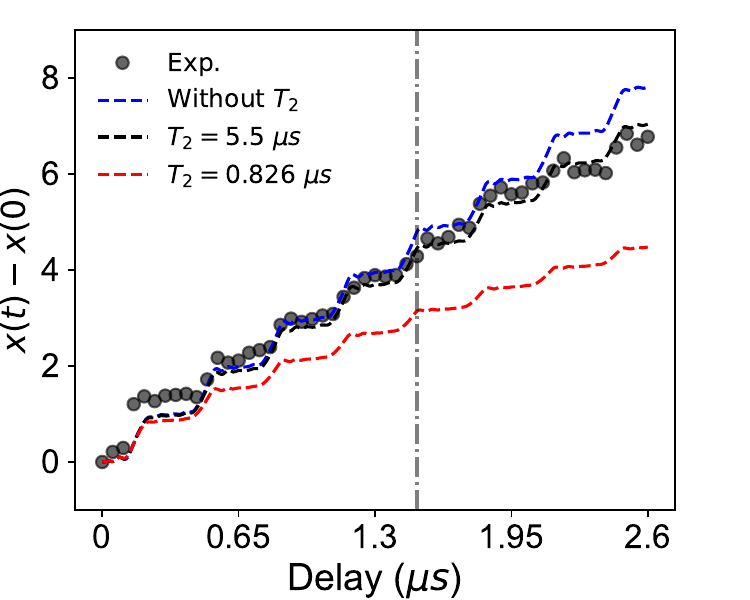}
	\caption{Experimental and numerical time evolution for topological pumping with different dephasing time $T_2$. The dark dots are experimental data, the blue, black and red curves are numerical simulations without dephasing, with $T_2=5.5$~$\mu$s and $T_2=0.826$~$\mu$s, respectively. }
	\label{fig:deltaxdephasing}
\end{figure}
The effects of decoherence, due to the interactions with the environment, are unavoidable in practical quantum simulations. In general, the influence on qubits can be classified into the energy relaxation effect, characterized by $T_1$, and the dephasing effect, identified by $T_{\varphi}$. As previously reported \cite{xu_emulating_2018, guo_observation_2021}, the effective dephasing time of a system with interactions is longer than the individually calibrated dephasing time $T_2^*$, due to the fact that eigenenergies of interacting systems depend weakly on each qubit flux.

Numerically, we calculate the topological pumping for the trajectory $\mathcal{C}_4$ introduced in the main text, with consideration of dephasing for a $12$-qubit chain, using the Lindblad master equation. For simplicity, we assume that all qubits have a uniform dephasing time to estimate the effective $T_{2}$ quantitatively. The direct comparison of numerical and experimental results is shown in Fig.~\ref{fig:deltaxdephasing}. We find that there is a good match between the experimental time evolution and the one obtained by numerical calculation when $T_2$ is chosen as $5.5$~$\mu$s. A deviation occurs when the evolution time beyond $3T$, due to the effect of short dephasing time. Moreover, we calibrated the states with a conserved particle number for the single-excitation or double-excitation initial states to mitigate the influence of energy relaxation. Overall, we conclude that our device can be approximately regarded as a closed quantum system when the experimental time within $1500$~ns.

\subsection{Pumping of double excitations}

\begin{figure}[t]
	\centering
	\includegraphics[width=0.97\linewidth]{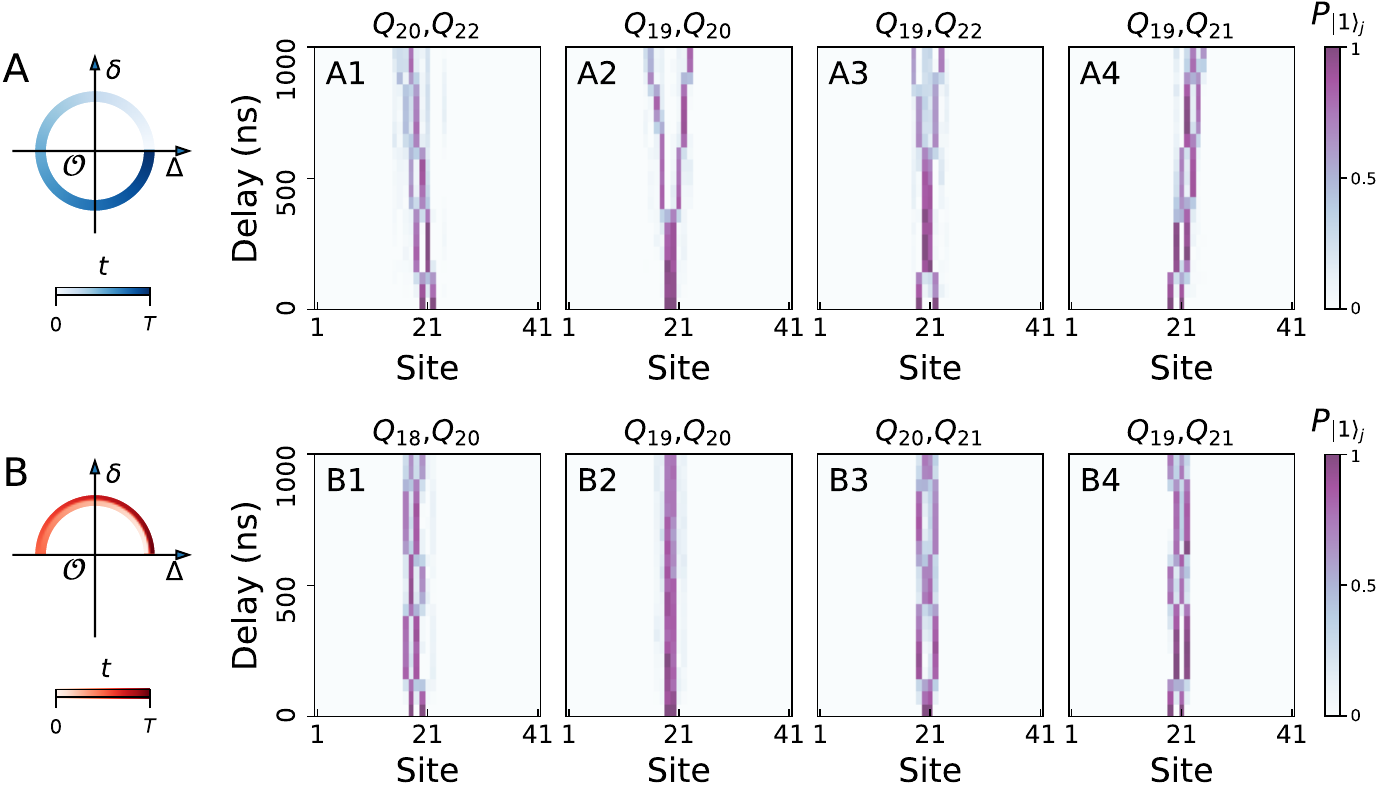}
	\caption{Thouless Pumping with two-photon excitation.
		(\textbf{A1}--\textbf{A4}) Experimental results of the time evolutions of the two-excitation states after initially exciting different qubits pairs for the trajectory in \textbf{A}.
		(\textbf{B1}--\textbf{B4}) Experimental results of the time evolutions of the two-excitation states by exciting different qubits pairs for the trajectory in \textbf{B}.}
	\label{fig:doublephoton}
\end{figure}

We also perform double-excitation experiments in the clean limit for two different trajectories. The trajectory as shown in Fig.~\ref{fig:doublephoton}A is $(\Delta_0\sin (2\pi t/T), \delta_0\cos(2\pi t/T))$, with $\Delta_0/2\pi=10$~MHz, $\delta_0/2\pi=2.5$~MHz, and $T=500$~ns, and the second trajectory as plotted in Fig.~\ref{fig:doublephoton}B is $(\Delta_0\cos (2\pi t /T), \delta_0|\sin (2\pi t/T)|)$ with the same parameters.

Since the initial excitations, prepared at odd and even sites, have opposite winding numbers in the RM model \eqref{eqRM}, $\Delta Q$ depends on both the trajectory and the parity combination of sites. In addition to the non-zero winding number, the condition that parity of the excited sites is the same, is required for topological nontrivial pumping. By exciting different qubits pairs, we obtain distant patterns of QWs as shown in Figs.~\ref{fig:doublephoton}A1--A4, and \ref{fig:doublephoton}B1--B4. Experimental results reveal that pumping of two-photon behaves like the transportation of two single-photons without interaction due to the fact that our system satisfies the hard-core limit \cite{yan_strongly_2019}.

\subsection{Pumping under on-site and hopping random disorder}
\begin{figure}[t]
	\centering
	\includegraphics[width=0.97\linewidth]{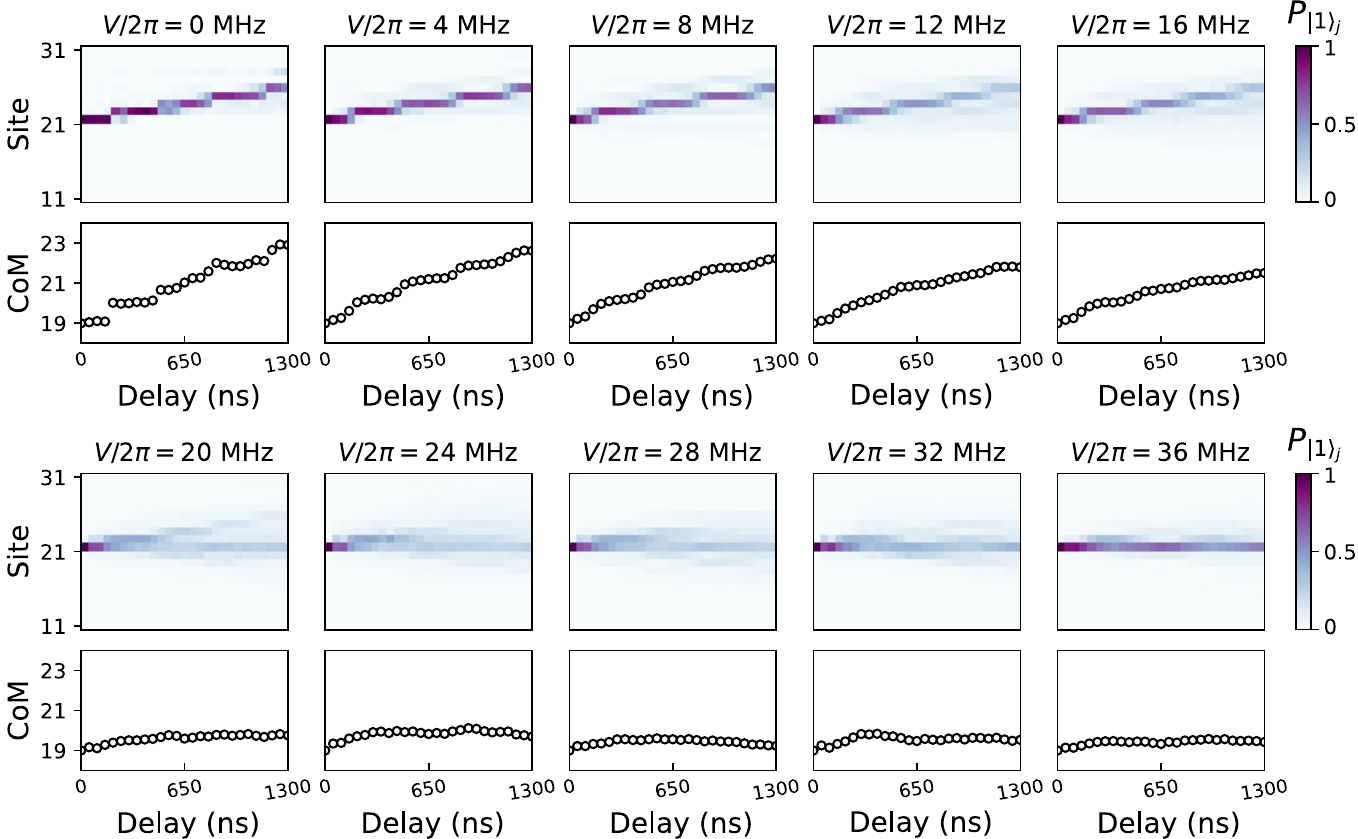}
	\caption{Experimental data of time evolutions of excitation probabilities and CoMs of pumping with on-site random disorder strengths ranging from $V/2\pi=0$~{MHz} to $36$~{MHz}. Topological pumping is robust against weak disorder strength and breaks down with a relatively strong disorder strength.}
	\label{fig:onsitedisorder}
\end{figure}

The experimental results of the single-photon excitation probabilities for the trajectory $\mathcal{C}_{\mathrm{on}}$ versus the on-site random disorder $V$ are plotted in Figs.~\ref{fig:onsitedisorder}.
The displacement of the CoM per cycle, $\delta x$, remains about two sites with weak disorder strength \SI{0}{\MHz} $\leq V/2\pi\lesssim$ \SI{8}{\MHz}.
The displacement of the CoM per cycle, $\delta x$, decreases to about \SI{0}{\MHz} with a strong disorder strength $V/2\pi\gtrsim$ \SI{28}{\MHz}.
Intuitively, the propagation with a considerably strong disorder is almost localized at the initial site, due to localization.

\begin{figure}[t]
	\centering
	\includegraphics[width=0.97\linewidth]{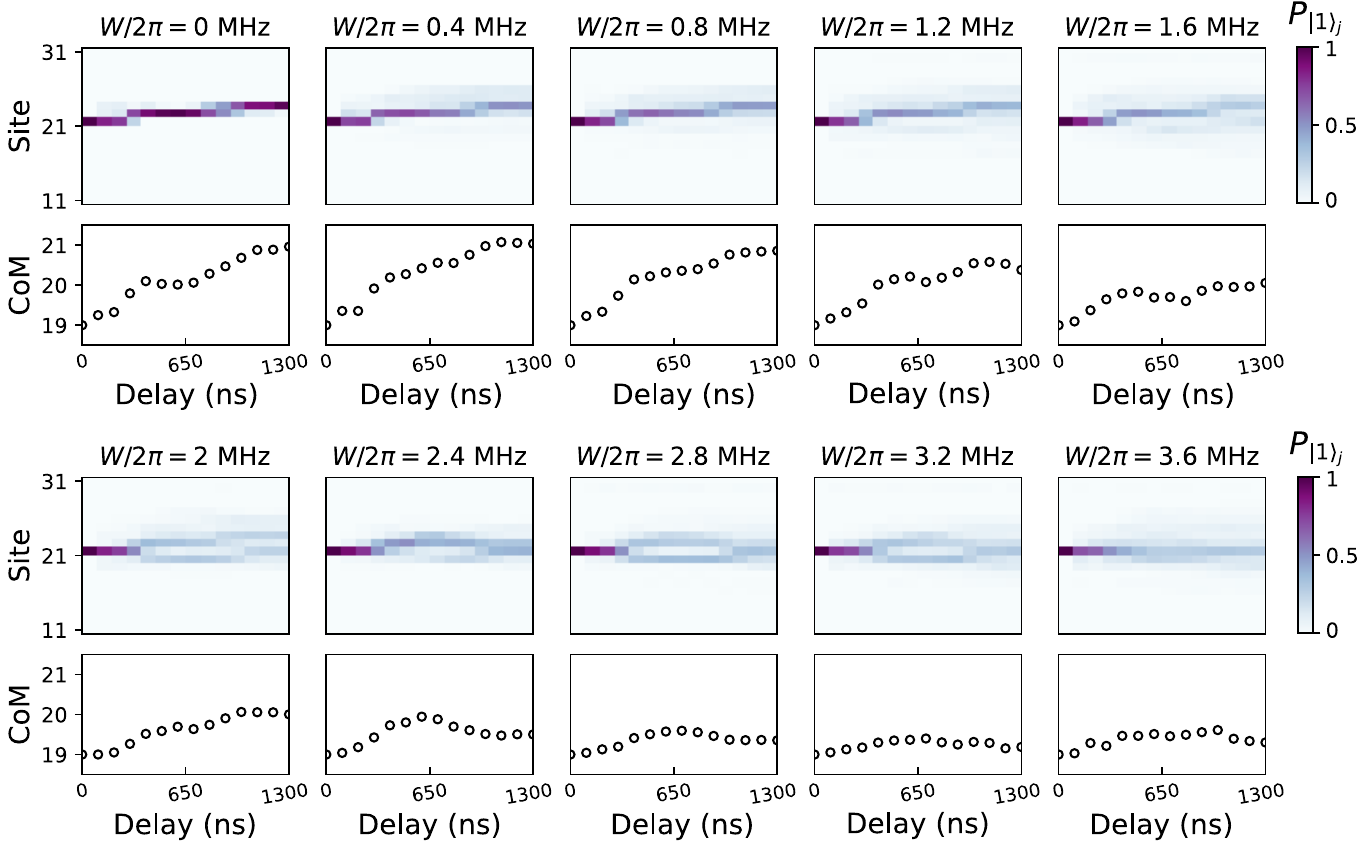}
	\caption{Experimental data of time evolutions of excitation probabilities and CoMs of pumping versus the hopping disorder strength ranging from $W/2\pi=0$~MHz to $3.6$~MHz. Similar as  pumping with on-site random disorder, the quantized transport is robust against weak disorder and breaks down with strong disorder.}
	\label{fig:hoppingdisorder}
\end{figure}

The experimental results of the single-photon excitation probabilities for the trajectory $\mathcal{C}_{\mathrm{hop}}$ versus the hopping random disorder strength $W$ are shown in Figs.~\ref{fig:hoppingdisorder}.
Similar to pumping with on-site random disorder, $\delta x$ persists about $2$ even with a weak disorder  strength and drops with a relatively strong disorder strength, as discussed in the Sec.~\ref{sec:disorder}.
More specifically, $\delta x\approx 2$ when \SI{0}{\MHz} $\leq W/2\pi\lesssim$ \SI{0.8}{\MHz}, and $\delta x\approx 0$ when $W/2\pi\gtrsim$ \SI{2.8}{\MHz}.

Therefore, \textit{we experimentally observe a competition between topology and disorder in Thouless pumping. With both on-site and hopping disorder, the quantized charge transport} (or change of polarization), indicated by the Chern number, \textit{is robust against disorder, for weak disorder, and breaks down when the disorder strength becomes relatively strong}.

\subsection{Pumping induced by on-site random disorder}
\begin{figure}[t]
	\centering
	\includegraphics[width=0.7\linewidth]{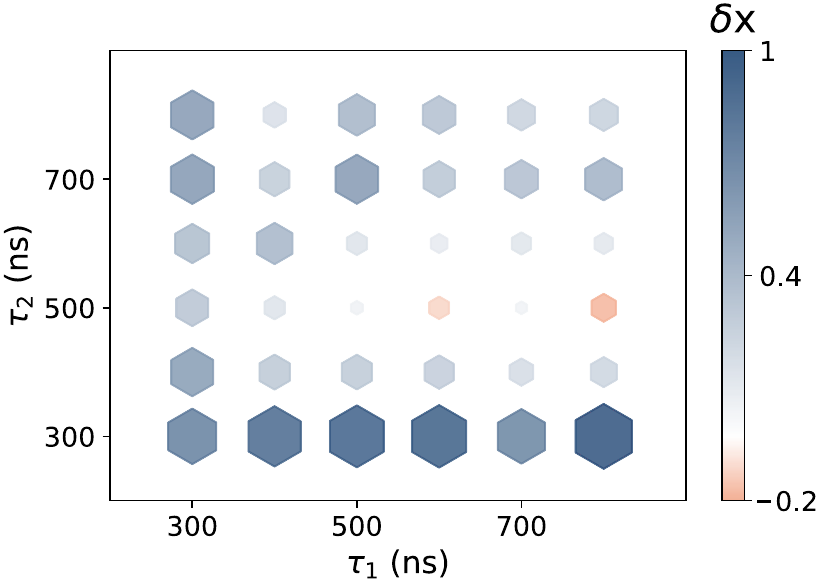}
	\caption{Experimental data of the displacement of CoM per cycle, $\delta x$ for pumping via the double-loop trajectory $\mathcal{C}_{\mathrm{dl}}$ against different periods $\tau_1$ and $\tau_2$.
		The results indicate that no quantized transport of two opposite-direction loops occurs with equal periods of two loops  $\tau_1=\tau_2=500$~ns.}
	\label{fig:onsitedicpperiod}
\end{figure}

\begin{figure}[t]
	\centering
	\includegraphics[width=0.97\linewidth]{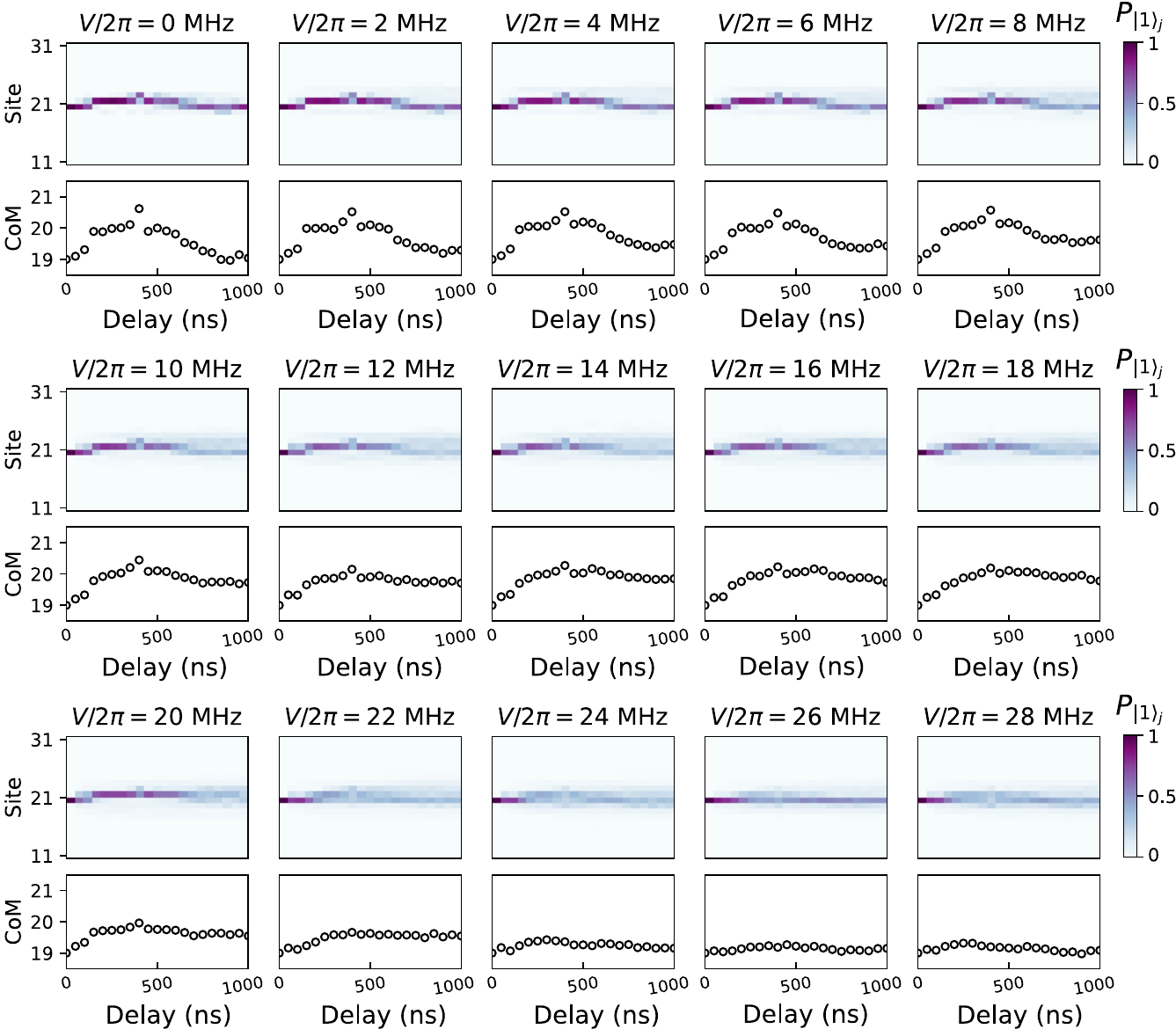}
	\caption{Experimental data of time evolutions of excitation probabilities and CoMs of pumping for the double-loop trajectory $\mathcal{C}_{\mathrm{dl}}$ versus the on-site random disorder strength $V$.
		The pumped amount $\Delta Q$ approximates $0$ in the clean limit, but  \emph{quantum  transport is observed with a moderate disorder strength and disappears again when disorder is strong enough}.}
	\label{fig:onsitedicp}
\end{figure}

By engineering the RM model Hamiltonian parameterized with $(\Delta, \delta)$ as shown in Eqs.~\eqref{eqDelta} and \eqref{eqdelta}, we can  study pumping for the double-loop trajectory $\mathcal{C}_{\mathrm{dl}}$.
First, we show $\delta x$ versus the periods of two individual loop, $\tau_1$ and $\tau_2$, to determine the appropriate evolving time. The numerical results, as plotted in Fig.~\ref{fig:onsitedicpperiod}, indicate that $\delta x$ is almost $0$ when $\tau_1=\tau_2=500$~ns, meaning the absence of topological pumping. Then, we show $\delta x$ versus the on-site random disorder strength $V$.
The results shown in Fig.~\ref{fig:onsitedicp} demonstrate topological transport occurs with a moderate disorder strength in the range \SI{5}{\MHz} $\lesssim V/2\pi\lesssim$ \SI{20}{\MHz}, indicating that topological pumping can be induced by on-site random disorder, and $\delta x$ drops to zero again when the disorder strength increases up to $V/2\pi\gtrsim$ \SI{25}{\MHz}.

\subsection{Pumping induced by quasi-periodic  hopping disorder}
\begin{figure}[t]
	\centering
	\includegraphics[width=0.97\linewidth]{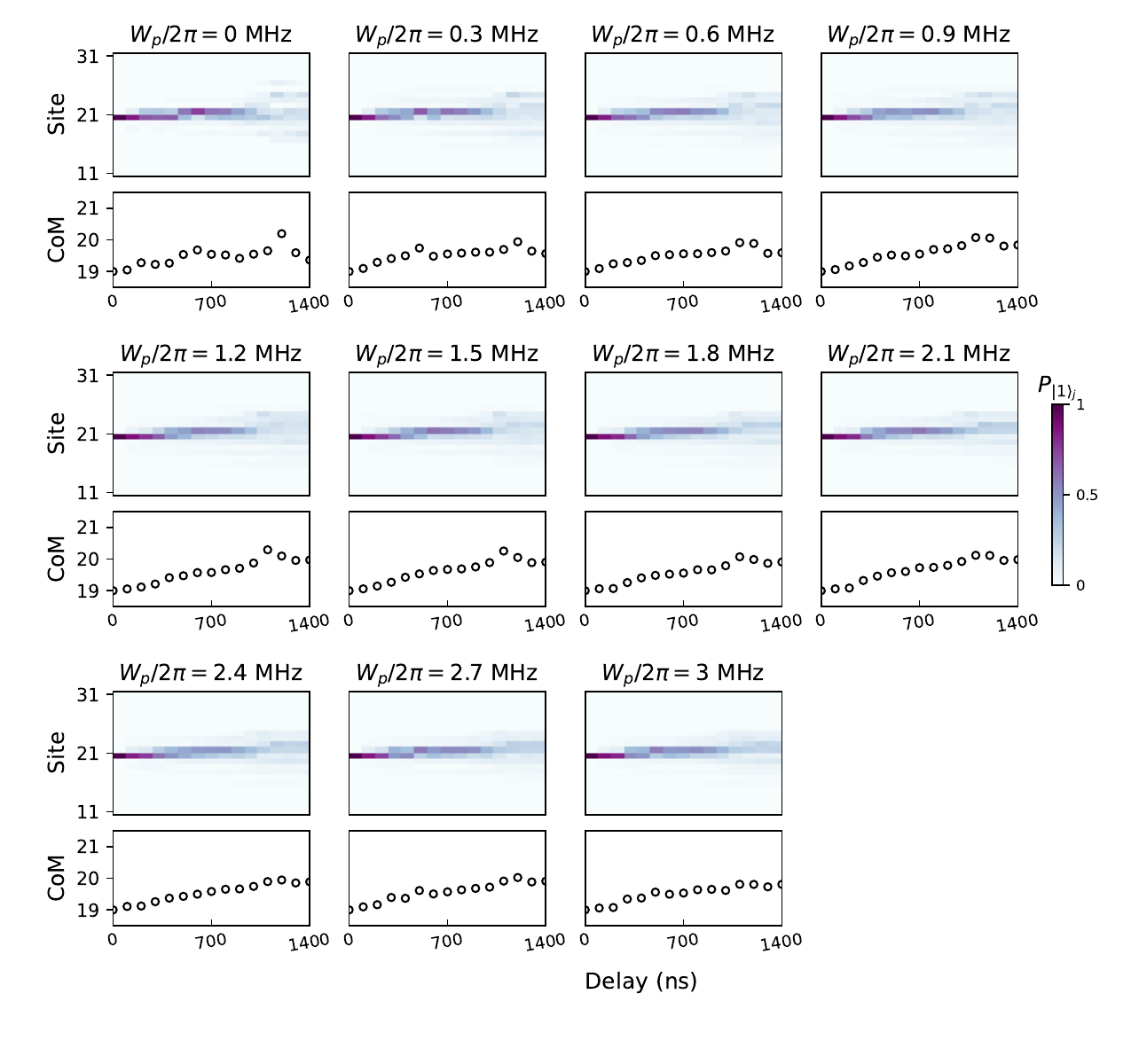}
	\caption{Experimental data of time evolutions of excitation probabilities and CoMs of pumping for single-loop trajectory $\mathcal{C}_{\mathrm{sl}}$ versus the disorder strength $W_p$.
		As the disorder strength increases,  CoM moves as the delay increase, which \emph{implies the existence of topological pumping induced by the quasi-periodic hopping  disorder} \cite{wu_quantized_2022}.}
	\label{fig:hoppingtatp}
\end{figure}

Quantized pumping is absent in the presence of random hopping disorder for the trajectory $\mathcal{C}_{\mathrm{sl}}$, as plotted in Fig.~\ref{fig:simulationhtatpchern}A. However, topological pumping may
occur when applying
quasi-periodic hopping disorder as discussed in Sec.~\ref{section:sl}.
We show $\delta x$ versus the quasi-periodic intracell hopping disorder strength $W_p$ in Fig.~\ref{fig:hoppingtatp}. Although $\delta x\approx 0.4$ is still nonzero even in the clean limit due to the non-adiabatic effect, the rising of $\delta x$ is captured as $W_p$ increases. In addition,
$\delta x> 0.9$ is observed when \SI{1.2}{\MHz} $\lesssim W_p/2\pi\lesssim$ \SI{2.7}{\MHz}, \textit{implying the existence of Thouless pumping induced by quasi-periodic disorder}.  In our experiments, the experimental scanning range of $W_p/2\pi$ is set up to $3$~MHz due to the limitation of Floquet engineering in our device.

\clearpage

\clearpage

\begin{tabular}[c]{|p{3cm}||p{3cm}|p{3cm}|}
	\hline
	\multicolumn{3}{|c|}{Experimental Data Table for Fig.~\ref{fig:onsitedisorder}} \\
	\hline
	$V/2\pi$ (MHz)& $\delta x$ & standard error\\
	\hline
	0 & 1.9563 & 0\\
	4 & 1.8107& 0.0337\\
	8 & 1.6136& 0.0461\\
	12 & 1.3985& 0.1025\\
	16 & 1.2552& 0.135\\
	20 & 0.3771& 0.1591\\
	24 & 0.347& 0.1388\\
	28 & 0.117& 0.1215\\
	32 & 0.2668& 0.1077\\
	36 & 0.2116& 0.1018\\
	\hline
\end{tabular}

\begin{tabular}{|p{3cm}||p{3cm}|p{3cm}|}
	\hline
	\multicolumn{3}{|c|}{Experimental Data Table for Fig.~\ref{fig:hoppingdisorder}} \\
	\hline
	$W/2\pi$ (MHz)& $\delta x$ & standard error\\
	\hline
	0 & 1.9618 & 0.0\\
	0.4 & 2.0358& 0.0638\\
	0.8 & 1.8591& 0.1424\\
	1.2 & 1.3807& 0.205\\
	1.6 & 1.0548& 0.1720\\
	2.0 & 1.0032& 0.1651\\
	2.4 & 0.5021& 0.0871\\
	2.8 & 0.3550& 0.1803\\
	3.2 & 0.1889& 0.2364\\
	3.6 & 0.3014& 0.2462\\
	\hline
\end{tabular}

\clearpage
\begin{tabular}[c]{|p{3cm}||p{3cm}|p{3cm}|}
	\hline
	\multicolumn{3}{|c|}{Experimental Data Table for Fig.~\ref{fig:onsitedicp}} \\
	\hline
	$V/2\pi$ (MHz)& $\delta x$ & standard error\\
	\hline
	0 & 0.0406 & 0.0\\
	2 & 0.2924 & 0.04995\\
	4 & 0.4992 & 0.0575\\
	6 & 0.5061 & 0.0814\\
	8 & 0.6636 & 0.0907\\
	10 & 0.7923 & 0.1037\\
	12 & 0.7073 & 0.0683\\
	14 & 0.8553 & 0.087\\
	16 & 0.7256 & 0.1049\\
	18 & 0.8147 & 0.1044\\
	20 & 0.6323 & 0.0967\\
	22 & 0.5447 & 0.1505\\
	24 & 0.1554 & 0.1144\\
	26 & 0.1501 & 0.1072\\
	28 & 0.0888 & 0.1227\\
	\hline
\end{tabular}

~\\
~\\
~\\

\begin{tabular}{|p{3cm}||p{3cm}|p{3cm}|}
	\hline
	\multicolumn{3}{|c|}{Experimental Data Table for Fig.~\ref{fig:hoppingtatp}} \\
	\hline
	$W_p/2\pi$ (MHz)& $\delta x$ & standard error\\
	\hline
	0.0 & 0.3632 & 0.0\\
	0.3 & 0.5683 & 0.0737\\
	0.6 & 0.5936 & 0.115\\
	0.9 & 0.8392 & 0.1007\\
	1.2 & 0.9743 & 0.1674\\
	1.5 & 0.9062 & 0.1531\\
	1.8 & 0.9075 & 0.1585\\
	2.1 & 0.9852 & 0.1538\\
	2.4 & 0.8888 & 0.1569\\
	2.7 & 0.91 & 0.0935\\
	3.0 & 0.8095 & 0.1171\\
	\hline
\end{tabular}

\clearpage
\bibliography{sm.bib}